\documentclass[a4paper,11pt]{article}
\usepackage{jinstpub}
\usepackage{inputenc,graphicx}
\usepackage{siunitx}
\usepackage{multirow}
\usepackage{makecell}
\usepackage{comment}
\usepackage[table]{xcolor}

\usepackage{lineno}

\newcommand{\pdensity}{$P_{\it{density}}$}
\newcommand{\maxflu}{$1 \times 10^{16}$ n$_{\text{eq}}$/cm$^2$}

\title{Testbeam Characterization of a SiGe BiCMOS Monolithic Silicon Pixel Detector with Internal Gain Layer}
\author[a,1]{L. Paolozzi\note{Corresponding author.},}
\author[a]{M. Milanesio,}
\author[a]{T. Moretti,}
\author[a]{R. Cardella,}
\author[a]{T. Kugathasan,}
\author[a]{A. Picardi,}
\author[b]{M. Elviretti,}
\author[b]{H. Rücker,}
\author[a]{F. Cadoux,}
\author[a]{R. Cardarelli,}
\author[a]{L. Cecconi,}
\author[a]{S. Débieux,}
\author[a]{Y. Favre,}
\author[a]{C. A. Fenoglio,}
\author[a]{D. Ferrere,}
\author[a]{S. Gonzalez-Sevilla,}
\author[a]{L. Iodice,}
\author[a,b]{R. Kotitsa,}
\author[a]{C. Magliocca,}
\author[a,c]{M. Nessi,}
\author[a]{A. Pizarro-Medina,}
\author[a]{J. Saidi,}
\author[a]{M. Vicente Barreto Pinto,}
\author[a]{S. Zambito,}
\author[a]{and G. Iacobucci}

\affiliation[a]{Département de Physique Nucléaire et Corpusculaire (DPNC), University of Geneva, 24 Quai Ernest-Ansermet, CH-1211 Geneva 4, Switzerland}

\affiliation[b]{CERN, CH-1211 Geneva 23, Switzerland}

\affiliation[c]{IHP — Leibniz-Institut für innovative Mikroelektronik, Im Technologiepark 25, Frankfurt (Oder), Germany}

\affiliation[d]{High Energy Accelerator Research Organization, Oho 1-1, Tsukuba-shi, Ibaraki-ken, Japan}

\emailAdd{lorenzo.paolozzi@unige.ch}

\abstract{
A monolithic silicon pixel ASIC prototype, produced in 2024 as part of the Horizon 2020 MONOLITH ERC Advanced project, was tested with a 120 GeV/c pion beam. The ASIC features a matrix of hexagonal pixels with a 100 $\mu$m pitch, read by low-noise, high-speed front-end electronics built using 130 nm SiGe BiCMOS technology. It includes the PicoAD sensor, which employs a continuous, deep PN junction to generate avalanche gain. 
Data were taken across power densities from 0.05 to 2.6 W/cm$^2$ and sensor bias voltages from 90 to 180 V. 
At the highest bias voltage, corresponding to an electron gain of 50, and maximum power density, an efficiency of (99.99 $\pm$ 0.01)\% was achieved. 
The time resolution at this working point was (24.3 $\pm$ 0.2) ps before time-walk correction, improving to (12.1 $\pm$ 0.3) ps after correction.
}
\begin{document}
\flushbottom
\maketitle

\abstract{something}

\section{Introduction}
\label{sec:Introduction}

Monolithic Active Pixel Sensors (MAPS \cite{peric}), which integrate the sensor and electronics within the same CMOS substrate, are becoming more and more common in particle physics. Despite their relatively complex design phase, MAPS provide all the benefits of standard CMOS processing while avoiding the production challenges and high costs associated with bump-bonded hybrid pixel sensors. A large surface of MAPS was recently installed in the ALICE experiment at the LHC \cite{alice} and is operated successfully. 

In the framework of the MONOLITH H2020 ERC Advanced project~\cite{monolith},
this group is developing very fast MAPS. Prototypes were produced using the commercial SG13G2 130 nm silicon-germanium (SiGe) BiCMOS process \cite{SG13G2} by IHP Microelectronics, which includes a very fast and low-noise SiGe Heterojunction Bipolar Transistor (HBT)  with a peak transition frequency $f_t$ =  350 GHz. 
In a testbeam experiment with minimum-ionizing particles (MIPs), a  prototype with a standard PN junction sensor without an internal gain layer
produced with this technology
provided full efficiency and time resolution of 20 ps~\cite{Zambito_2023}. 
Several of these prototypes were irradiated with 70 MeV protons up to a fluence \maxflu~ and tested with radioactive sources~\cite{Milanesio_2024} and in a testbeam experiment with MIPs~\cite{Moretti_2024}. The SiGe HBT front-end electronics not only survived such large proton fluence but was still able to recover full efficiency and provide a time resolution of 45 ps with only a modest increase in sensor bias voltage.
This performance underscores SiGe HBT radiation tolerance, meeting the stringent requirements of the challenging High-Luminosity LHC environment.

In parallel to the development of front-end electronics, the MONOLITH ERC project is also exploiting the picosecond avalanche detector (PicoAD)~\cite{PicoADpatent}, a novel detector that implements a continuous deep gain layer to boost the timing performance. 
A proof-of-concept monolithic PicoAD was produced in 2022 using the photolithography masks of a prototype able to provide 36 ps~\cite{Iacobucci:2021ukp} with a sensor without internal gain layer. 
The proof-of-concept PicoAD was operated at a gain of approximately 20~\cite{picoad_gain}, obtaining full efficiency and an average time resolution of 17 ps~\cite{PicoAD_TB}. The time resolution was found to depend strongly on the hit position on the pixel area, which varied between 13 ps at the center of the pixel and 25 ps in the inter-pixel region~\cite{PicoAD_TB}.

Recently, new PicoAD prototypes were produced using the photolithography masks of the ASIC that was characterized intensively in \cite{Zambito_2023, milanesio2023radiation, Milanesio_2024}.
This paper presents the results of an experiment at the CERN SPS testbeam facility with the latest PicoAD prototype, in which both the time resolution and the efficiency were measured with 120 GeV/c pions.
\section{Description of the ASIC}
\label{sec:ASICDescription}

\begin{figure}[!htb]
\centering
\includegraphics[width=.33\textwidth]{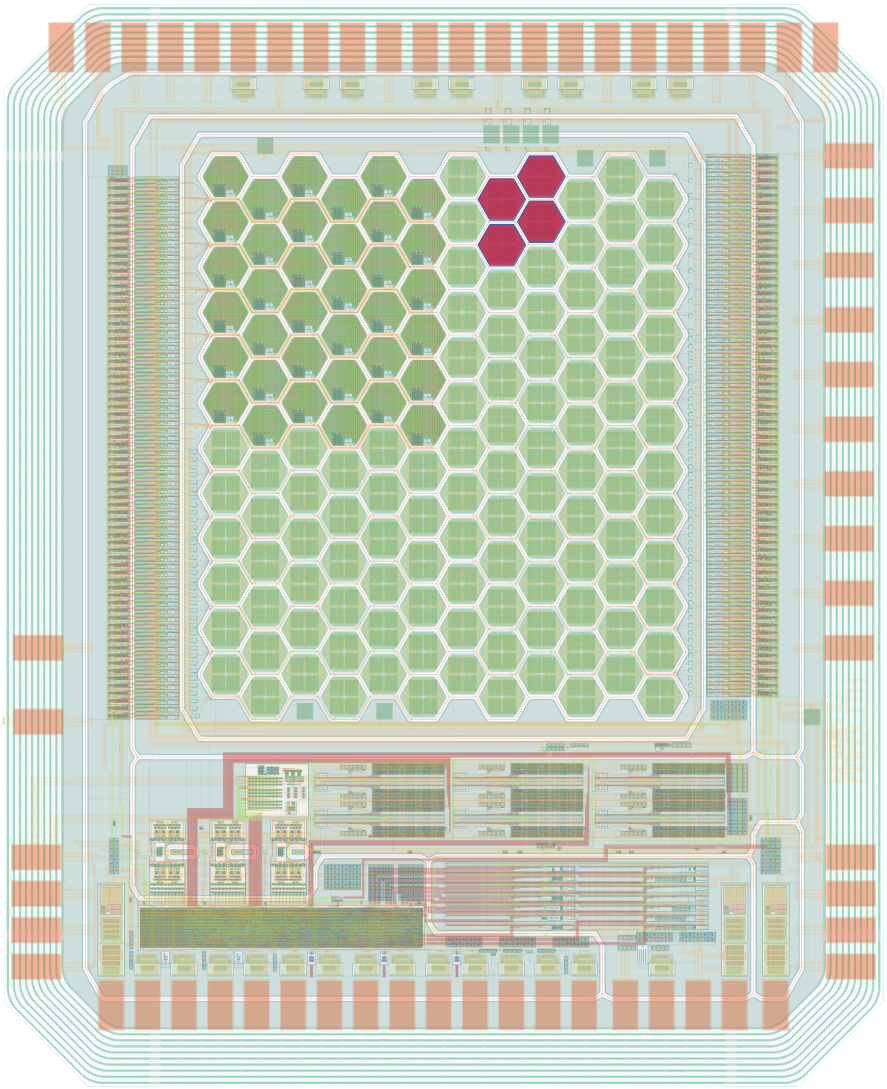}
\raisebox{0.37\height}{\includegraphics[width=.55\textwidth]{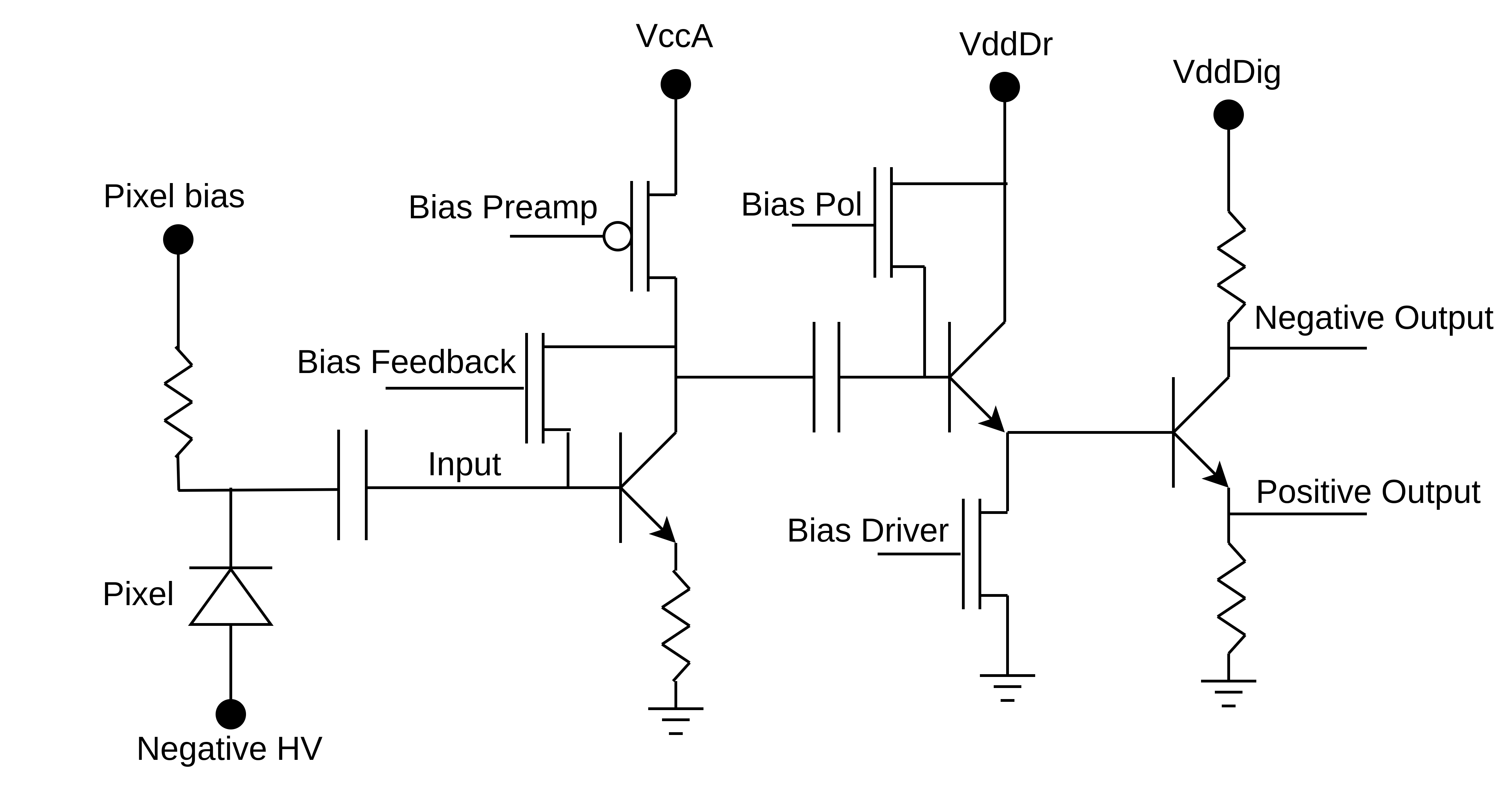}}
\caption{\label{fig:layout} (Left) Layout view of the 2022 prototype ASIC. The four pixels in red are the analog pixels whose amplifier's output is connected directly to an analog driver with differential output. (Right) Schematic view of the front-end electronics and driver configuration of the analog pixels.}
\end{figure}

The ASIC studied in this paper contains a matrix of 12$\times$12 hexagonal pixels of 65 $\mu$m side, corresponding to a pitch of approximately 100 $\mu$m, produced by IHP in the SG132G2 130 nm SiGe BiCMOS technology. 
It is an evolution of the previous prototype~\cite{Iacobucci:2021ukp}, with improved front-end electronics. The ASIC contains four analog pixels (highlighted in red in Figure \ref{fig:layout}) in which the amplifier is connected to an analog driver to be read by an oscilloscope.  The right panel of Figure \ref{fig:layout} shows a schematic view of the front-end electronics and analog driver.

\section{PicoAD Production}
\label{sec:PicoADProduction}

\begin{figure}[!htb]
\centering
\includegraphics[width=.65\textwidth]{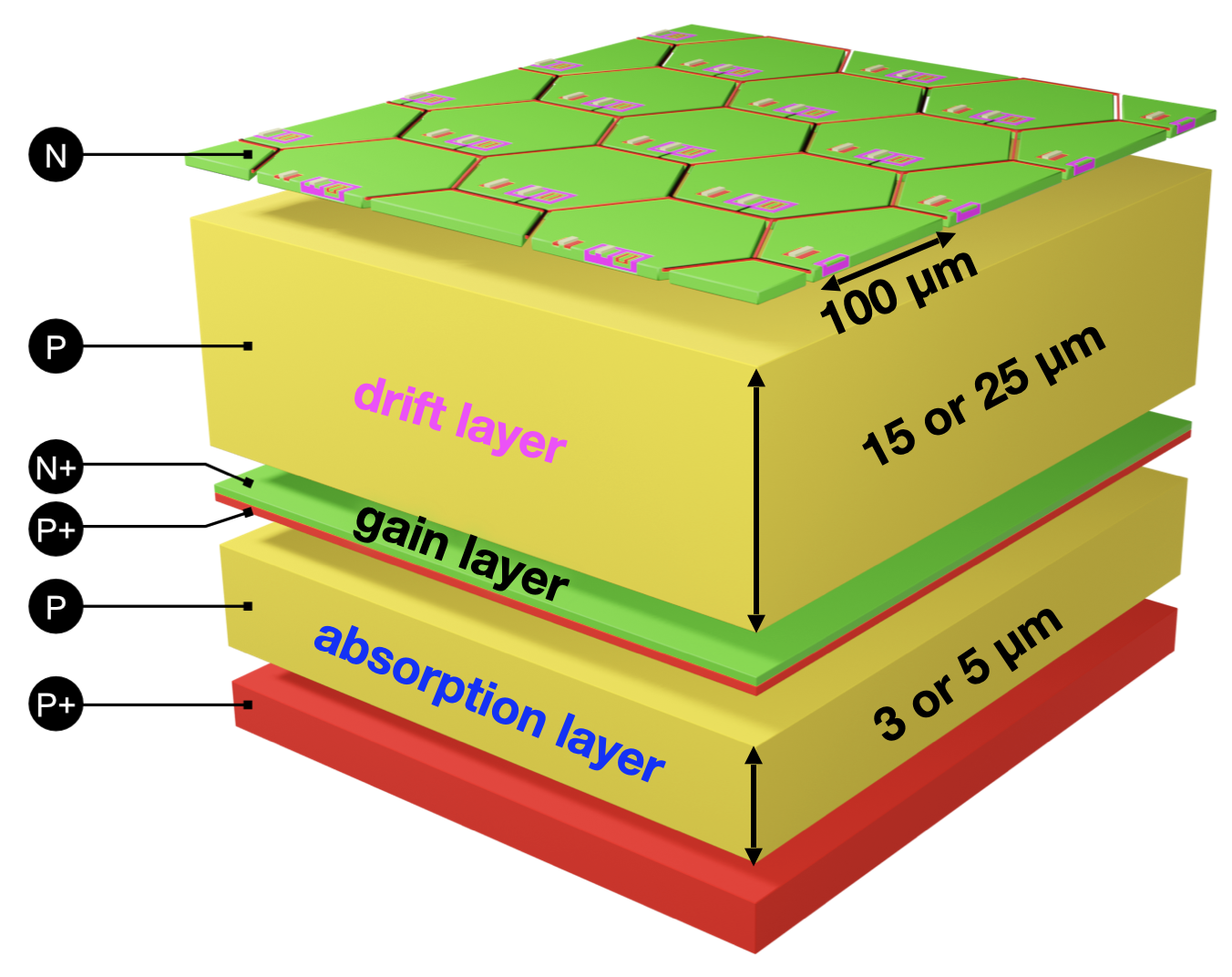}
\caption{\label{fig:PicoADlayout} Layout view of the PicoAD prototype with hexagonal pixels with 65 $\mu$m side. Prototypes were produced in several flavours: four different gain-layer implant doses, thickness of the absorption layer of 3 and 5 $\mu$m, and thickness of the drift layer of 15 and 25 $\mu$m. All the results presented in this paper were obtained with the prototype with the largest implant dose, 5 $\mu$m absorption layer and  15 $\mu$m drift layer. 
}
\end{figure}

The PicoAD sensor has a multi-junction structure to engineer the electric field inside the silicon active area. The aim is to provide an electric field large enough to generate impact ionization and produce avalanche gain multiplication only of the electrons generated in a very thin zone ("absorption layer" in Figure~\ref{fig:PicoADlayout}), thus minimising the charge-collection (Landau) noise.

The monolithic  ASIC prototypes were manufactured by the Leibniz Institute for High Performance Microelectronics (IHP). 
Prior to CMOS processing, the PicoAD wafers were manufactured
following the procedure explained in detail in~\cite{picoad_gain}, that requires the following steps (from bottom to top in Figure~\ref{fig:PicoADlayout}):
\begin{itemize}
    \item first, an epitaxial  layer of 3 or 5 $\mu$m was grown on a low resistivity silicon wafer; the primary electrons produced in this thin absorption layer will cross the gain layer and be multiplied, producing the fast component of the signal with minimal Landau noise; 
    \item then a gain layer was implanted, with several ion doses determined with a TCAD study;
    \item finally, a second epitaxial layer of 15 or 25 $\mu$m (called "drift layer") was grown
   , which allows drifting of the primary electrons produced in the drift layer and the secondary electrons produced in the absorption layer toward the collection electrode (on the top of Figure~\ref{fig:PicoADlayout}).
\end{itemize}
The wafers were then CMOS processed to produce the ultra-fast and low-noise electronics of~\cite{Zambito_2023}. 

A total of 15  different PicoAD flavors were produced with the two absorption and the two drift layer thicknesses and with different boron and arsenic implant doses of the gain layer. The testbeam data presented in this paper were taken with  ASICs that have a PicoAD sensor with 5 $\mu$m-thick absorption layer, 15 $\mu$m-thick drift layer, and the highest dose implanted.
It was found that, at temperature T = $-5^{\circ}$C, a bias voltage of 70 V was sufficient to fully deplete the sensors, while silicon breakdown  occurred at 200 V. 

Prior to the testbeam, data were collected at the same temperature using  $^{55}$Fe and $^{90}$Sr radioactive sources.
These data were used to calculate the electron gain at various bias voltages and to convert the threshold values from mV to electrons. The results are summarized in Table~\ref{tab:maintable}.

\section{Testbeam Measurements}
\label{sec:TestbeamMeasurements}

\subsection{Experimental Setup}
\label{subsec:ExperimentalSetup}
\begin{figure}[htb!]
\centering 
\includegraphics[width=.8\textwidth]{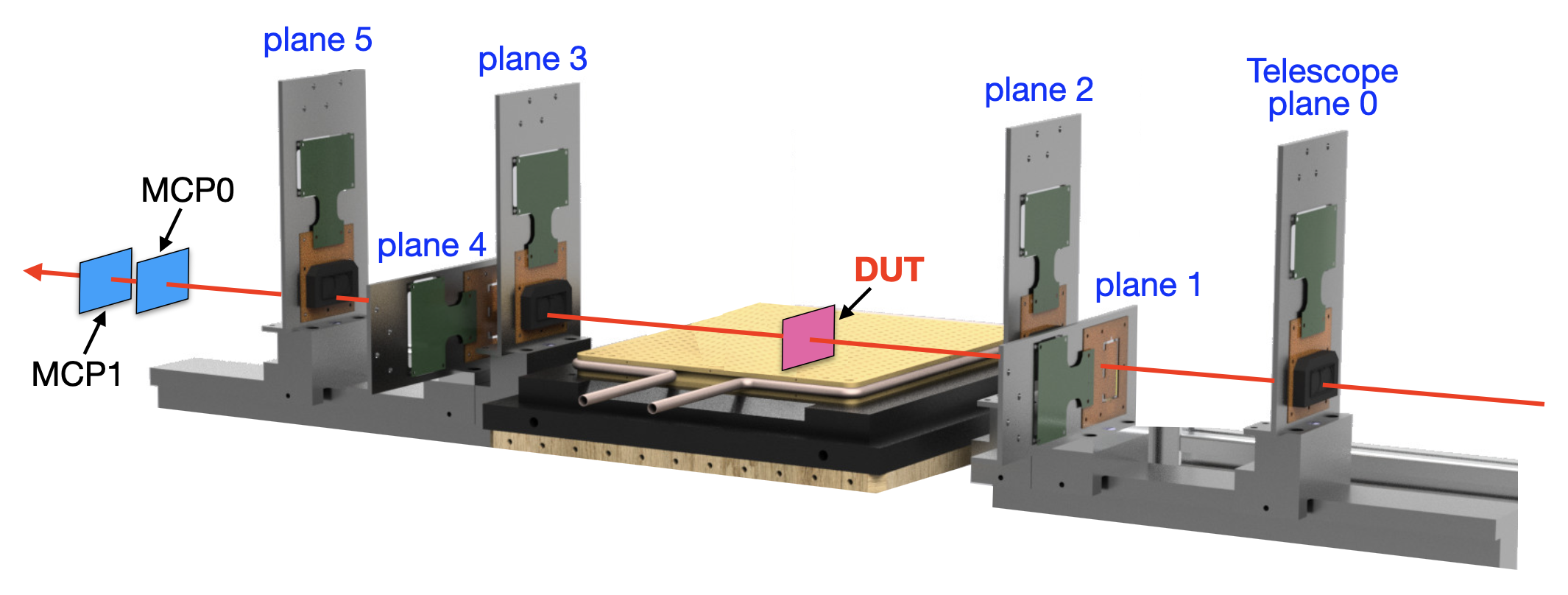}
\caption{\label{fig:FEI4} Schematic view of the experimental setup, showing the six planes of the UNIGE FE-I4 telescope, the DUT, and the two MCPs. The red arrow shows the direction of the pion beam.}
\end{figure}

Data were taken at the CERN SPS testbeam facility using pions with a momentum of 120 GeV/c.
Figure \ref{fig:FEI4} shows a schematic view of the experimental setup. It comprised:
\begin{itemize}
    \item the monolithic silicon ASIC device under test (DUT);
    \item the UNIGE FE-I4 testbeam telescope \cite{FEI4_telescope} to reconstruct the tracks of the pions and give the trigger to the two oscilloscopes;
    \item two Photonis PP2365AC micro-channel plate photomultiplier tubes (called MCP0 and MCP1) to provide a precise reference time, which were mounted outside the FE-I4 telescope downstream of the pion beam; 
    \item an oscilloscope Lecroy WaveMaster 820 Zi-A (40 GS/s and set to 4 GHz analog bandwidth); its four channels were used to read the two polarities of the signal from one of the analog channels of the DUT as well as the two MCPs;
    \item an oscilloscope Lecroy WaveRunner 9404 (20 GS/s and set to 4 GHz analog bandwidth); its four channels were used to read the positive polarities of the remaining three analog pixels of the DUT and to check the shape of the trigger given by the telescope.
\end{itemize}
The DUT was placed in the middle of the six planes of the telescope in a cooling box and was operated through all the data acquisition at a temperature of -5 $^{\circ}$C. A region of interest of 250 $\times$ 250 $\mu m^{2}$ centered around the four analog pixels of the DUT was imposed on the first telescope plane: the coincidence between this region and the last plane of the telescope gave the trigger to the system.

For each triggered event, the waveforms of the DUT and the two MCPs were acquired by the oscilloscope in a time window of 500 ns, with the DUT and MPCs signals arriving at the center of the window. The first 200 ns, which did not contain the signals, were used to measure the baseline of the voltage noise and its standard deviation $\sigma_{V}$.

\subsection{Data Samples}
\label{subsec:Data Samples}
During the testbeam experiment, 
data were acquired at different working conditions of the ASIC, varying the sensor bias voltage and the power density of the preamplifier. 
The minimum value of the sensor bias voltage was HV~=~90~V, corresponding to an electron gain of 4. The maximum value was 180 V, which is about 20 V before the sensor breakdown, corresponding to an electron gain of 50.
The front-end electronics was operated with absorbed currents between 3 and 160 $\mu A$, corresponding to power densities between 0.05 and 2.6 W/cm$^2$. 

Table~\ref{tab:maintable} lists the data samples collected. For each operating point, a few million tracks were triggered in the region of interest defined by the testbeam telescope and recorded. 
The offline selection criteria to single out an unbiased sample of 
events 
included:
\begin{itemize}
\item only one reconstructed telescope track per event, which was required to have:
\begin{itemize}
    \item a recorded hit in all the six planes of the telescope
    \item a reconstructed $\chi^{2}/NDF$ < 1.6
\end{itemize}
\item one and only one hit in both the MCP0 and the MCP1;
only events with signals in both 
MCP0 and MCP1 with amplitudes not saturated by the oscilloscope readout scale were retained, so to allow for accurate time-walk correction and timing reference. 
\end{itemize}

These criteria reduced the data samples to the number of selected events reported in Table~\ref{tab:maintable}.

\begin{figure}[htb!]
\centering 
\includegraphics[width=.60\textwidth]{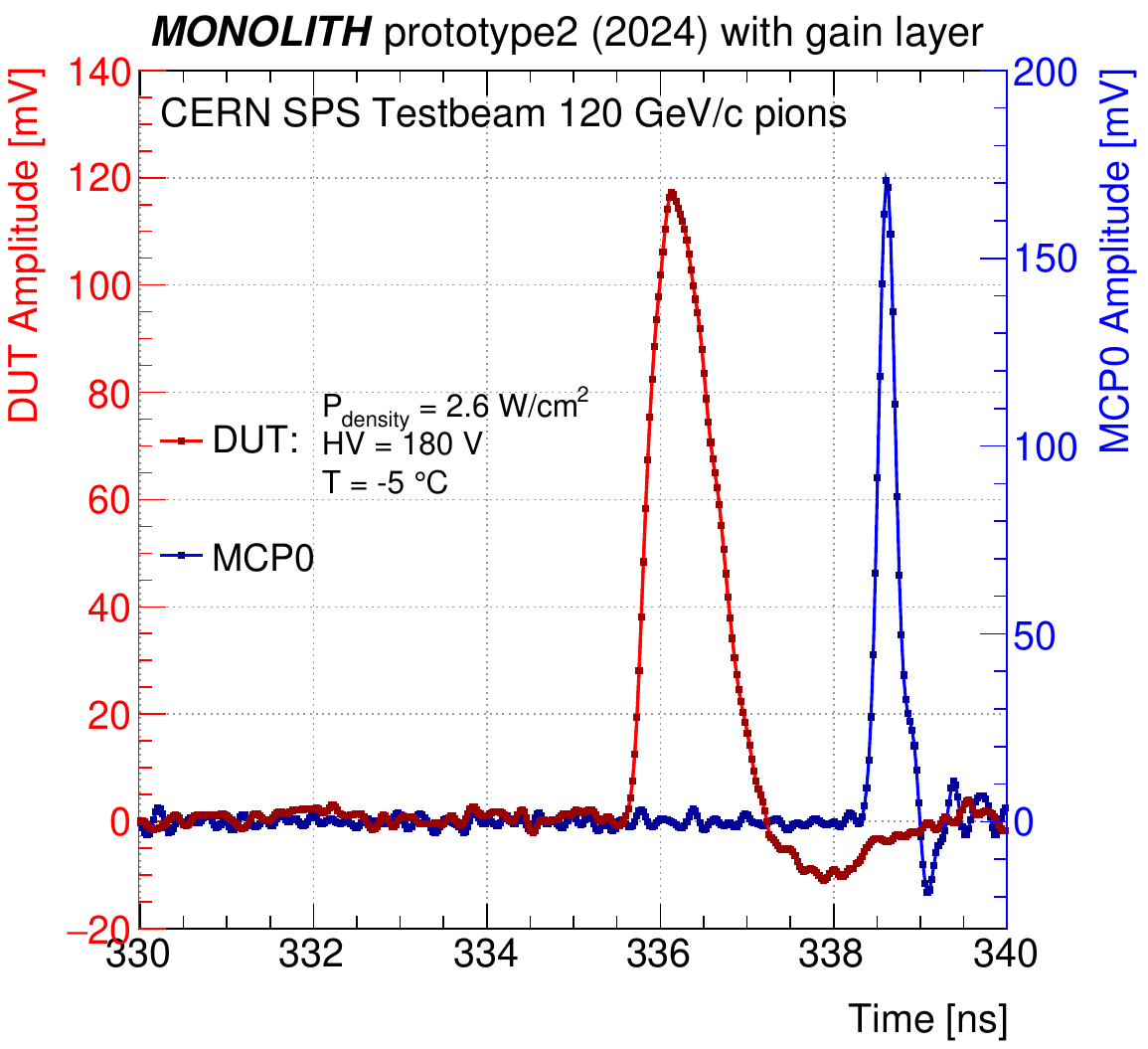}
\caption{\label{fig:waveformExample} 
Example of waveforms acquired by the oscilloscope in a 10 ns time interval around the DUT signal (dots interpolated with red segments) and the MCP0 signal (dots interpolated with blue segments). The dots represent the amplitude values acquired with 25 ps time binning, while the segments connecting the dots show the linear interpolation that was used during the data analysis.
}
\end{figure}

Figure \ref{fig:waveformExample} shows an example of waveforms of the DUT and the MCP0. 

\begin{figure}[htb!]
\centering 
\includegraphics[width=.60\textwidth]{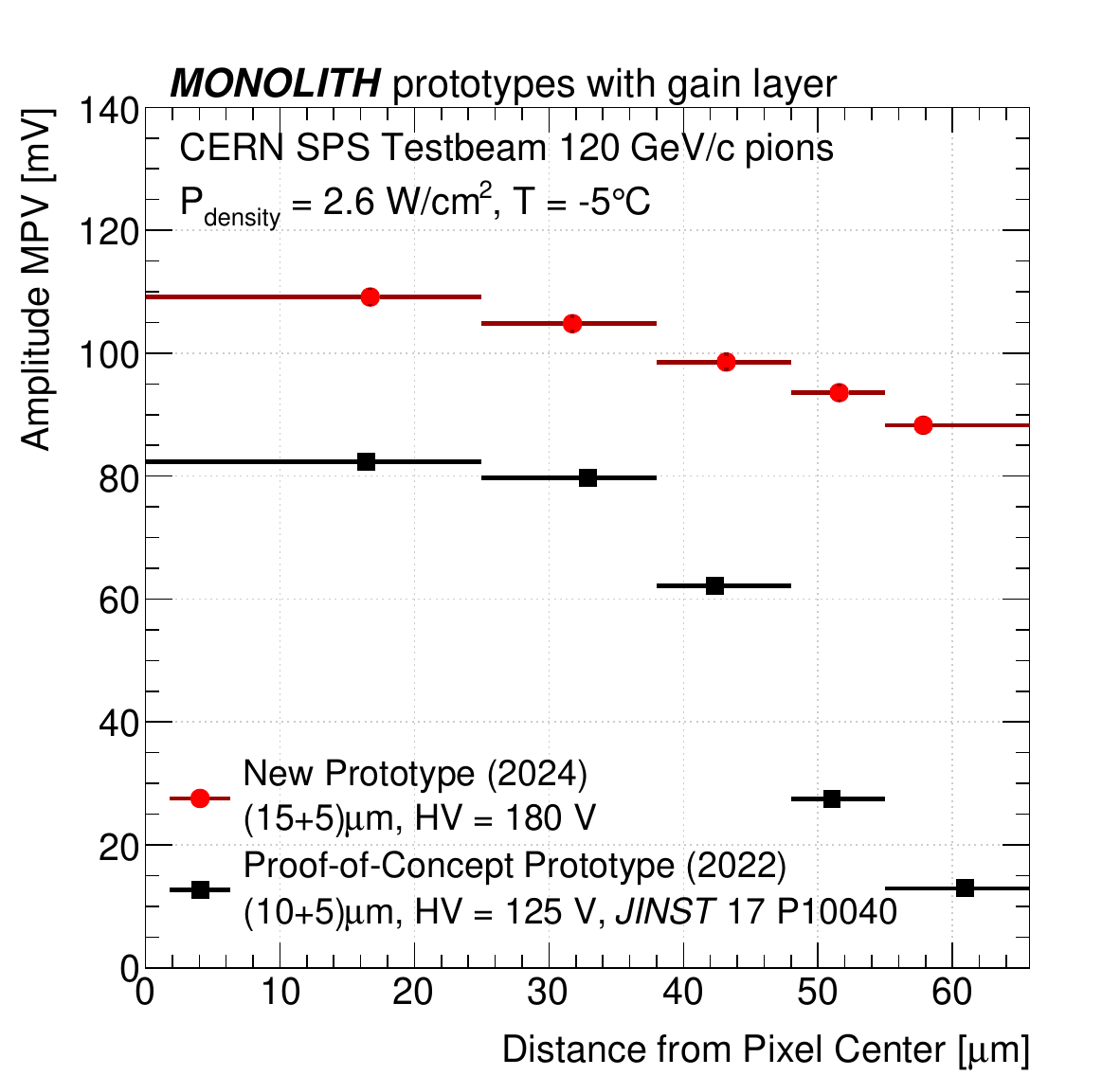}
\caption{\label{fig:ampRadius}
Comparison of the MPV of the signal amplitude as a function of the distance from the pixel center for the current prototype (red dots) and the proof-of-concept prototype (black squares).}
\end{figure}

\begin{table}[]
\centering
\renewcommand{\arraystretch}{1.2}%
\begin{tabular}{|c|c|c|c|c|c|}
\hline

\vspace{-3pt}
~~~~~HV~~~~~~   & ~~Electron~~  & ~~~~$P_{\it{density}}$~~~~    & ~Threshold~ & ~~~Selected~~~  & ~~Efficiency~~ \\
{[}V{]}     & Gain      & {[}W/cm$^2${]}        & [electrons] ([mV])   & Events    & {[}\%{]} \\
\hline
\multirow{5}{*}{90}& \multirow{5}{*}{4} 
                        & ~~~0.05   & 895 ~(3) & ~~8'499   & \cellcolor{black!10}$43.7^{~\!+1.5}_{-2.8}$ \\
                    &   & ~0.1      & 950 ~(4) & ~~4'191   & \cellcolor{black!10}$70.3^{~\!+1.0}_{-2.0}$ \\
                    &   & ~0.3      & 790 ~(5) & 12'666   & \cellcolor{black!10}$95.5^{~\!+0.3}_{-0.5}$ \\
                    &   & ~\!1.0    & 725 ~(7) & 11'022   & \cellcolor{black!10}$96.8^{~\!+0.3}_{-0.6}$ \\
                    &   & 2.6       & 840 ~(9) & 12'339   & \cellcolor{black!10}$98.4^{~\!+0.2}_{-0.4}$ \\
\hline
\multirow{5}{*}{120}& \multirow{5}{*}{10} 
                        & ~~~0.05   & 895 ~(3) & 11'406   & \cellcolor{black!10}$69.9^{~\!+0.4}_{-0.7}$ \\
                    &   & ~0.1      & 950 ~(4) & ~~8'199   & \cellcolor{black!10}$90.5^{~\!+0.4}_{-0.8}$ \\
                    &   & ~0.3      & 790 ~(5) & 10'425   & $99.1^{~\!+0.1}_{-0.3}$ \\
                    &   & ~\!1.0    & 725 ~(7) & ~~4'542   &$99.3^{~\!+0.2}_{-0.4}$ \\
                    &   & 2.6       & 840 ~(9) & 31'524    & $99.73^{~\!+0.04}_{-0.09}$ \\
\hline
\multirow{5}{*}{150}& \multirow{5}{*}{18} 
                        & ~~~0.05   & 895 ~(3) & 14'505   & \cellcolor{black!10}$93.7^{~\!+0.2}_{-0.4}$ \\
                    &   & ~0.1      & 950 ~(4) & ~~9'558   & $99.1^{~\!+0.1}_{-0.3}$ \\
                    &   & ~0.3      & 790 ~(5) & 34'416    & $99.65^{~\!+0.03}_{-0.07}$ \\
                    &   & ~\!1.0    & 725 ~(7) & ~~9'567   & $99.77^{~\!+0.04}_{-0.12}$ \\
                    &   & 2.6       & 840 ~(9) & 44'430    & $99.98^{~\!+0.01}_{-0.03}$ \\
\hline
\multirow{3}{*}{160}& \multirow{3}{*}{21} 
                        & ~0.3      & 790 ~(5) & 16'290   & $99.84^{~\!+0.02}_{-0.05}$ \\
                    &   & ~\!1.0    & 725 ~(7) & ~~9'009   & $99.83^{~\!+0.03}_{-0.07}$ \\
                    &   & 2.6       & 840 ~(9) & ~~7'704   & $99.97^{~\!+0.02}_{-0.06}$ \\
\hline
\multirow{5}{*}{170}& \multirow{5}{*}{33} 
                        & ~~~0.05   & 895 ~(3) & 17'364    & $99.27^{~\!+0.07}_{-0.15}$ \\
                    &   & ~0.1      & 950 ~(4) & ~9'612   & $99.73^{~\!+0.07}_{-0.18}$ \\
                    &   & ~0.3      & 790 ~(5) & ~4'161   & $99.8^{~\!+0.1}_{-0.3}$ \\
                    &   & ~\!1.0    & 725 ~(7) & ~9'798   & $99.85^{~\!+0.07}_{-0.17}$ \\
                    &   & 2.6       & 840 ~(9) & 11'487   & $99.96^{~\!+0.04}_{-0.12}$ \\
\hline
\multirow{5}{*}{180}& \multirow{5}{*}{50}
                        & ~~~0.05   & 895 ~(3) & 13'083   & $99.76^{~\!+0.04}_{-0.09}$ \\
                    &   & ~0.1      & 950 ~(4) & 10'167   & $99.88^{~\!+0.04}_{-0.12}$ \\
                    &   & ~0.3      & 790 ~(5) & ~9'807   & $99.92^{~\!+0.03}_{-0.10}$ \\
                    &   & ~\!1.0    & 725 ~(7) & ~9'069   & $99.84^{~\!+0.05}_{-0.13}$ \\
                    &   & 2.6       & 840 ~(9) & 53'973    & $99.99^{~\!+0.01}_{-0.02}$ \\
\hline
\end{tabular}
\caption{List of the 28 datasets acquired during the testbeam experiment. The table reports the sensor bias voltage and the corresponding PicoAD electron gain, the \pdensity~at which the front-end electronics was operated, the threshold (in electrons and corresponding voltage) applied to the signal amplitude, the number of selected events according to the criteria of section~\ref{subsec:Data Samples}, and the detection efficiency. 
The gain and the threshold value in electrons were obtained with data taken with $^{55}$Fe and $^{90}$Sr radioactive sources.
The datasets with efficiency smaller than 99\%, which will not be considered for the time resolution  studies, are highlighted in grey.}
\label{tab:maintable}
\end{table}
For each dataset, the DUT signal amplitudes were distributed
in bins of distance of the hit position from the pixel center
and fitted with a Landau function, extracting the most probable value (MPV) of the Landau. Figure~\ref{fig:ampRadius} illustrates a comparison of the Landau MPV for the current prototype and the proof-of-concept version \cite{PicoAD_TB}. In the current design, a 20\% decrease in MPV is observed from the pixel center to the interpixel region, compared to a nearly eightfold reduction in the earlier version. This improvement results from two key sensor modifications: increasing the drift layer thickness from 10 to 15 $\mu$m and raising the active layer resistivity from 50 to 350 $\Omega$cm in the current prototype.

\subsection{Detection Efficiency}
\label{subsec:Efficiency}
\begin{figure}[htb!]
\centering 
\includegraphics[width=.49\textwidth]{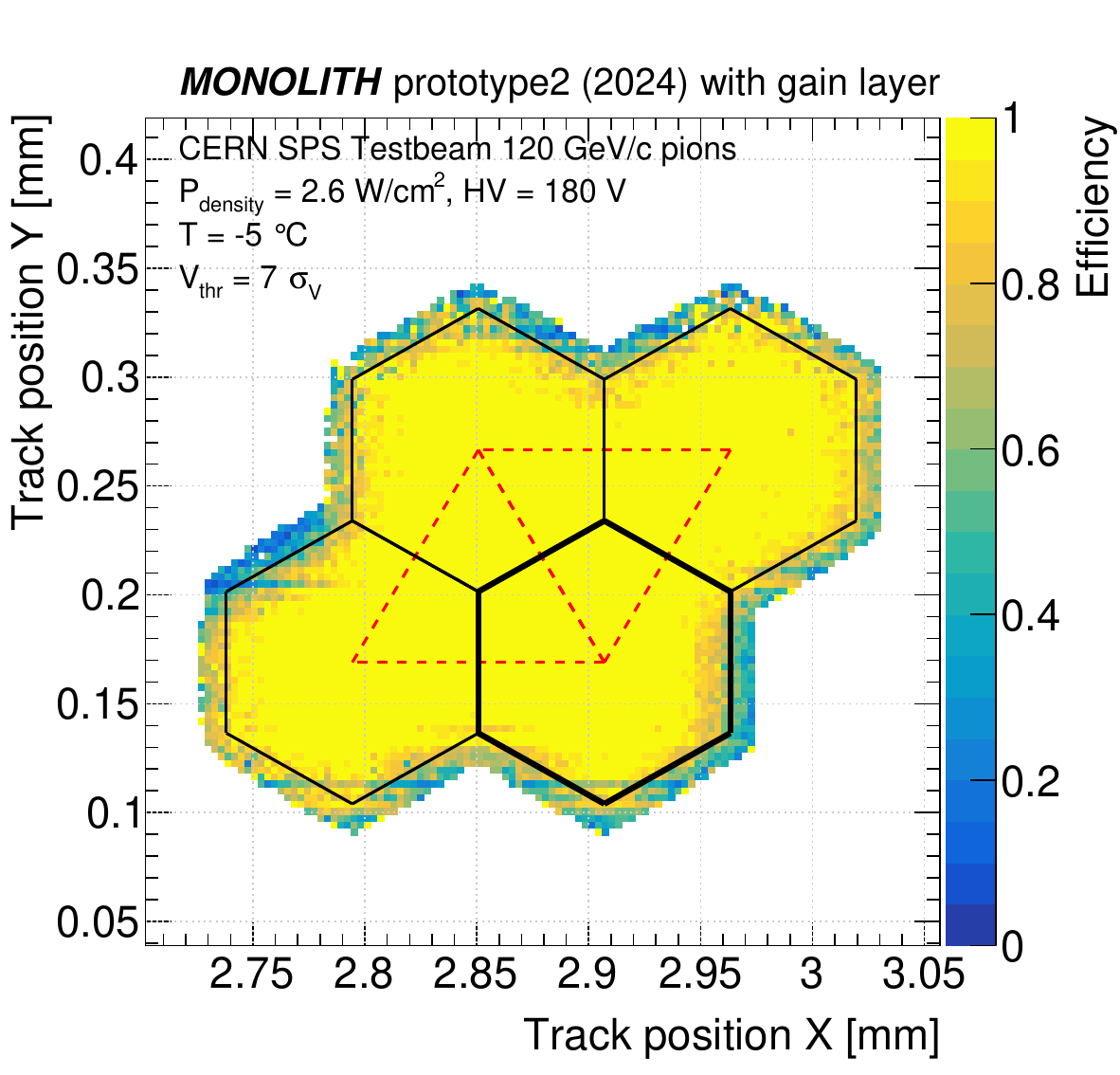}
~~\includegraphics[width=.49\textwidth]{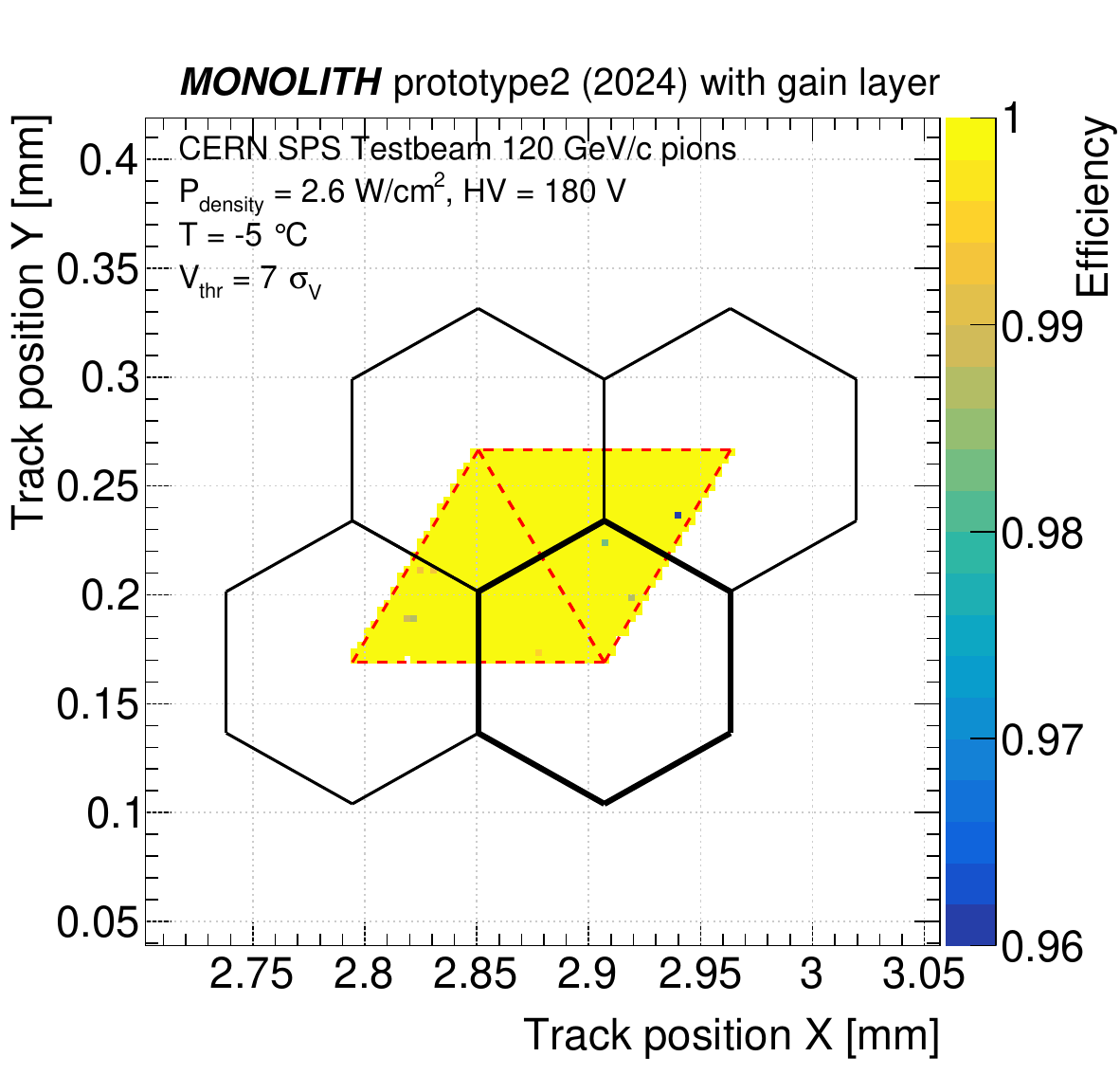}
\caption{\label{fig:efficiencyMaps} 
(Left) Map of the detection efficiency of the four analog pixels. The effect of the finite pointing resolution of the testbeam telescope is visible at the external edges of the pixels. (Right) Map of the efficiency restricted to the region of the two triangles obtained connecting the centers of the pixels, which is not biased by the telescope resolution. The results refer to the sample taken with HV = 180 V and \pdensity~= 2.6 W/cm$^2$. Only the bottom-right pixel (highlighted by thicker lines) was connected to the 25 ps sampling oscilloscope and thus used to compute the time resolution.}
\end{figure}

The detection efficiency was measured 
using the sample of all events that had the telescope track crossing the DUT plane within the area of the four analog pixels. The tracks extrapolated within 15 $\mu$m from the external edge of the surface of the four pixels were kept in the sample to account for the telescope's spatial resolution~\cite{FEI4_telescope}.  
An event was defined as efficient if it had a signal in the DUT with an amplitude larger than 7 times the voltage noise $\sigma_V$. 
The left panel of Figure~\ref{fig:efficiencyMaps} shows the efficiency map for the four analog pixels.
While the efficiency around the five internal inter-pixel regions remains constant and larger than 95\%, the efficiency is found to degrade in the 15 $\mu$m around the 14 external edges, confirming the influence of the telescope spatial resolution. 

To obtain an unbiased measurement of the efficiency, the sample of telescope tracks was restricted to those that crossed the DUT inside the two triangles obtained connecting the centers of the four analog pixels (shown by dashed red segments in Figure~\ref{fig:efficiencyMaps}).
The right panel of Figure~\ref{fig:efficiencyMaps} shows the efficiency map of this unbiased sample with a colour scale palette starting at 96\%.
The dataset shown in Figure~\ref{fig:efficiencyMaps}, acquired with the DUT operated at HV = 180 V and \pdensity~ = 2.6 W/cm$^2$,  gives an efficiency of $(99.99 \pm 0.01)$\%.
The Clopper-Pearson frequentist approach~\cite{clopper_pearson} with a confidence level of 95\% was used to estimate the uncertainties. 

\begin{figure}[htb!]
\centering 
\includegraphics[width=.8\textwidth]{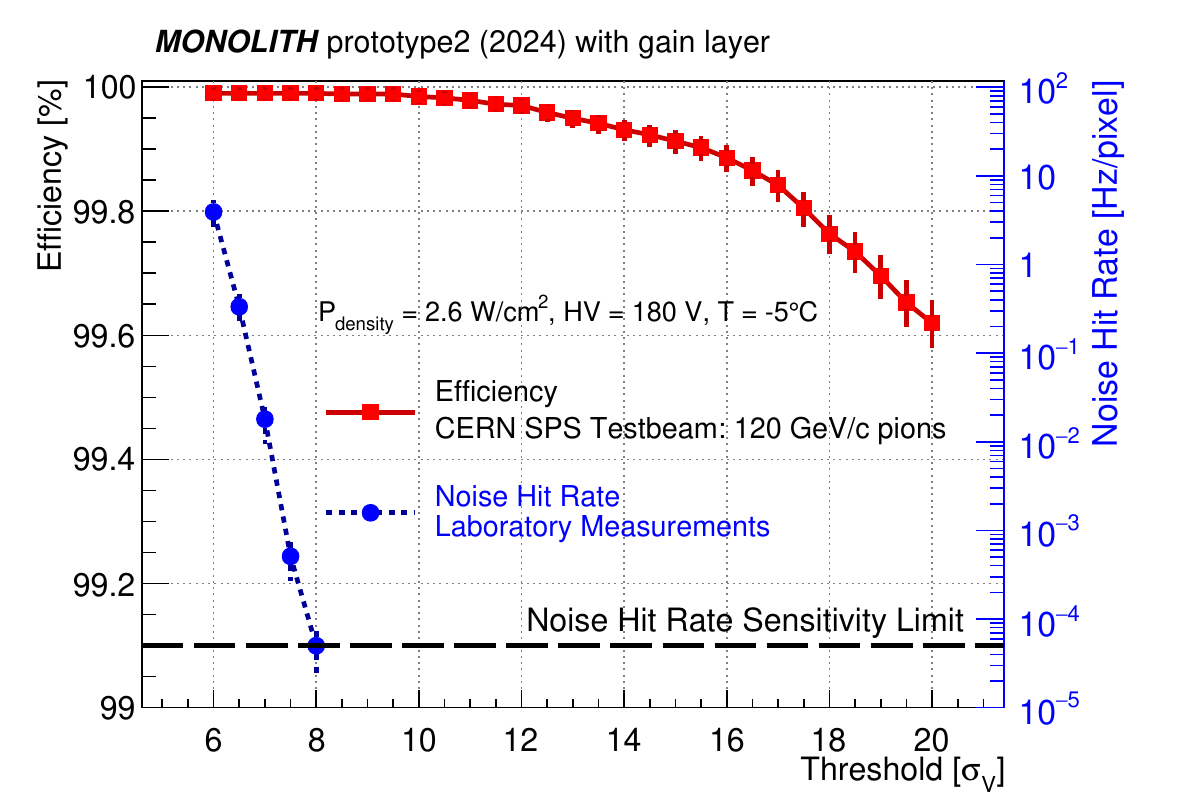}
\caption{\label{fig:noiseHitRate} 
Detection efficiency (red squares) and noise hit rate (blue dots) as a function of the threshold for the dataset taken with HV = 180 V and \pdensity~= 2.6 W/cm$^2$.
The threshold is given in units of the voltage noise $\sigma_V$.}
\end{figure}

Figure \ref{fig:noiseHitRate} shows the detection efficiency as a function of the threshold for the same dataset.
Remarkably, even for a threshold  20 times the voltage noise $\sigma_{V}$, the efficiency remains larger than 99.6\%.
Figure \ref{fig:noiseHitRate} also reports the noise hit rate. The threshold of 7$\sigma_{V}$ utilized throughout this paper is well within the efficiency plateau and corresponds to a noise hit rate per pixel of $2\times10^{-2}$ Hz. Variation of the threshold by one unit of $\sigma_V$ produces a change in noise hit rate of approximately two orders of magnitude.

\begin{figure}[htb!]
\centering 
\includegraphics[width=.6\textwidth]{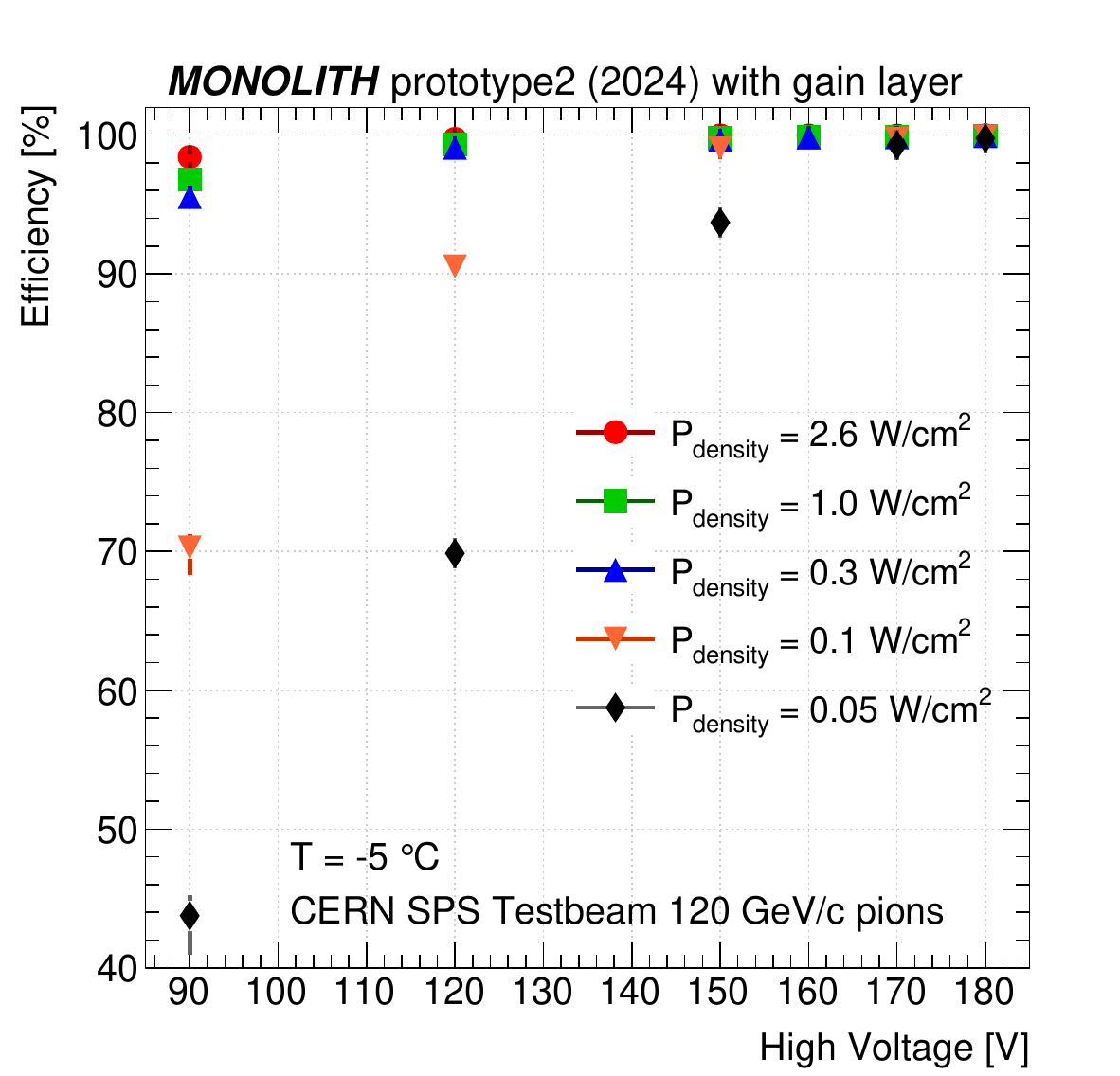}
\caption{\label{fig:effSummary} 
Detection efficiency as a function of the bias voltage of the sensor for five values of the power density of the preamplifier.
}
\end{figure}

The detection efficiency for all the datasets is reported in Table~\ref{tab:maintable} and in Figure~\ref{fig:effSummary}.
As expected, efficiency increases with the preamplifier's power density and with the sensor bias voltage. The former is due to an increase in the front-end gain, and the latter is due to a larger charge multiplication in the gain layer.
The measurements show that for \pdensity~$\ge$ 0.1 W/cm$^2$ the detector reaches the plateau of efficiency at HV = 150 V. At \pdensity~= 0.05 W/cm$^2$ a sensor bias voltage of 170 V is needed to obtain a detection efficiency larger than 99\%.

\begin{figure}[htb!]
\centering
\includegraphics[width=.6\textwidth]{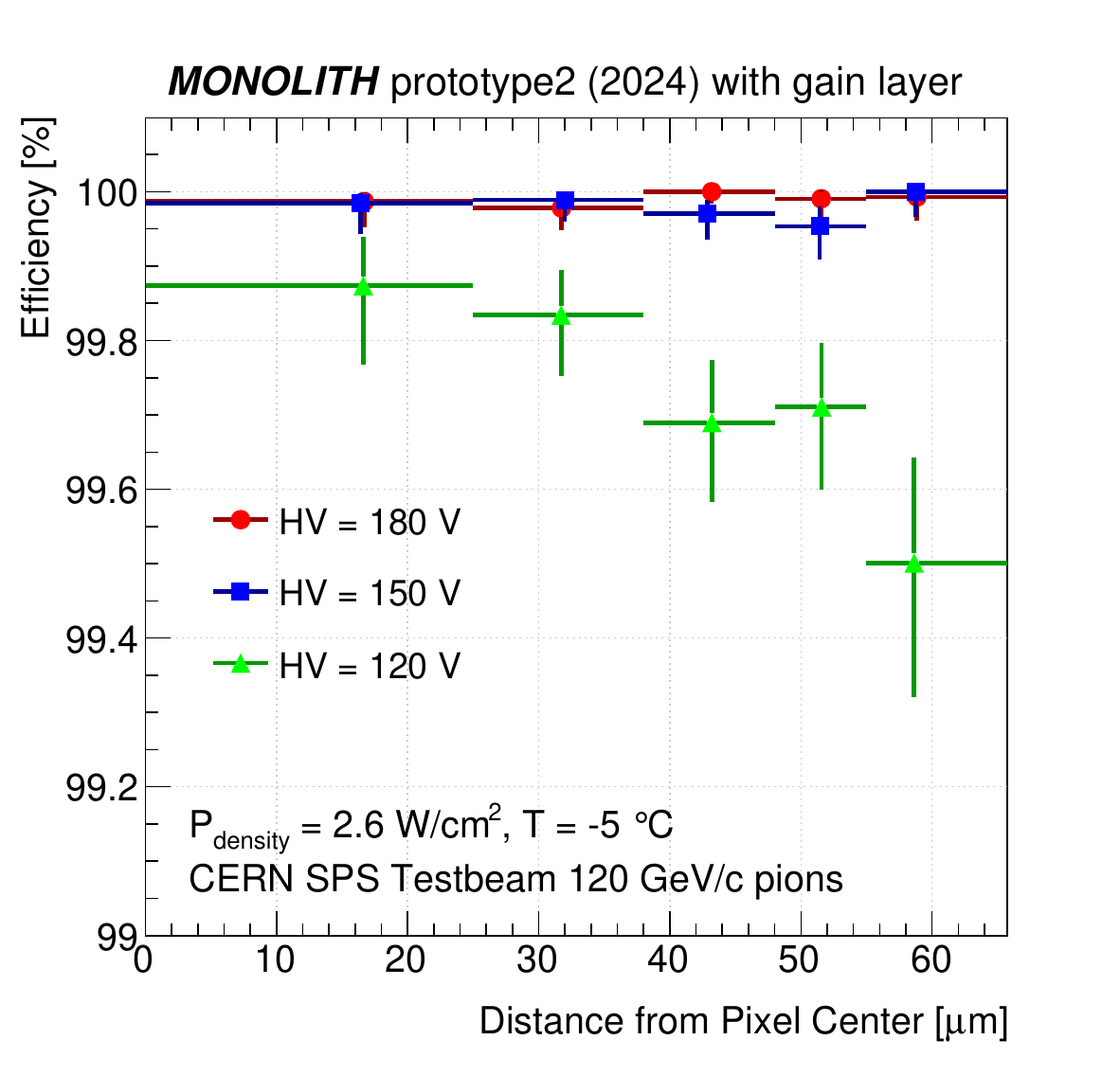}
\caption{\label{fig:eff_vs_distance}
Detection efficiency in bins of the distance of the hit position from the pixel center. The data show the results obtained with \pdensity~= 2.6 W/cm$^2$ for three values of the sensor bias voltage. In each bin, the data points are plotted at the mean value of the track distance from the pixel center.}
\end{figure}

Figure \ref{fig:eff_vs_distance} shows the detection efficiency in bins of the distance from the pixel center for the dataset taken with the highest \pdensity~= 2.6 W/cm$^2$.  The efficiency is well above 99.9\% and does not depend on the hit position for HV = 150 V and HV = 180 V. 
At HV = 120 V, the detection efficiency decreases to 99.5\% near the pixel edge, likely due to charge sharing between adjacent pixels.

\subsection{Time Resolution}
\label{subsec:TimeResolution}
The time resolution was measured starting from the same sample of events used for the measurement of the detection efficiency of Figure~\ref{fig:efficiencyMaps}. Since only the bottom-right analog pixel of the DUT in Figure~\ref{fig:efficiencyMaps} was connected to the oscilloscope sampling at 25 ps, only the subsample of hits in this analog pixel was considered for further analysis, which corresponds to approximately one-third of the events used for the measurement of the efficiency.
The time-resolution measurement was restricted to the datasets in Table~\ref{tab:maintable} with detection efficiency larger than 99\%.


The oscilloscope samplings were interpolated linearly to measure the time of arrival (ToA) of the signals.
In the case of the two MCPs, the ToA was taken as the time at which the interpolated signal crossed 50\% of the maximum value of the amplitude, thus correcting most of the time walk.
In the case of the DUT, the ToA was taken as
the time at which the interpolated signal crossed a fixed voltage threshold. 
The threshold value V$_{\text{ToA}}$ was set at the integer value in mV closest to seven times the voltage noise $\sigma_V$ measured in that dataset. 
As reported in Table~\ref{tab:maintable}, the threshold ranged from 3 to 9 mV, depending on the front-end electronics working point.  
Finally, for each DUT signal, the ToA value was determined by linearly interpolating between the two oscilloscope samplings with amplitudes just below and above the threshold.

The  dataset with the largest statistics was used to 
compute the time resolution of the MCP0
using the method adopted in \cite{Zambito_2023}:
the time resolution for each of the three ToA differences, constructed from pairs among DUT, MCP0, and MCP1, was measured by fitting the distribution with a Gaussian functional form, and the resulting system of three equations with three unknowns was solved.
This method yielded a time resolution for MCP0 of  $\sigma_{t,\text{MCP0}} = (5.4 \pm 0.4)$ ps. 

Once this value was established, the time resolution of the DUT at each operating point was determined by performing a Gaussian fit on the ToA$_{\text{DUT}}$ - ToA$_{\text{MCP0}}$ distribution and then subtracting the measured $\sigma_{t,\text{MCP0}}$ in quadrature.

\begin{figure}[htb!]
\centering
\includegraphics[width=.32\textwidth]{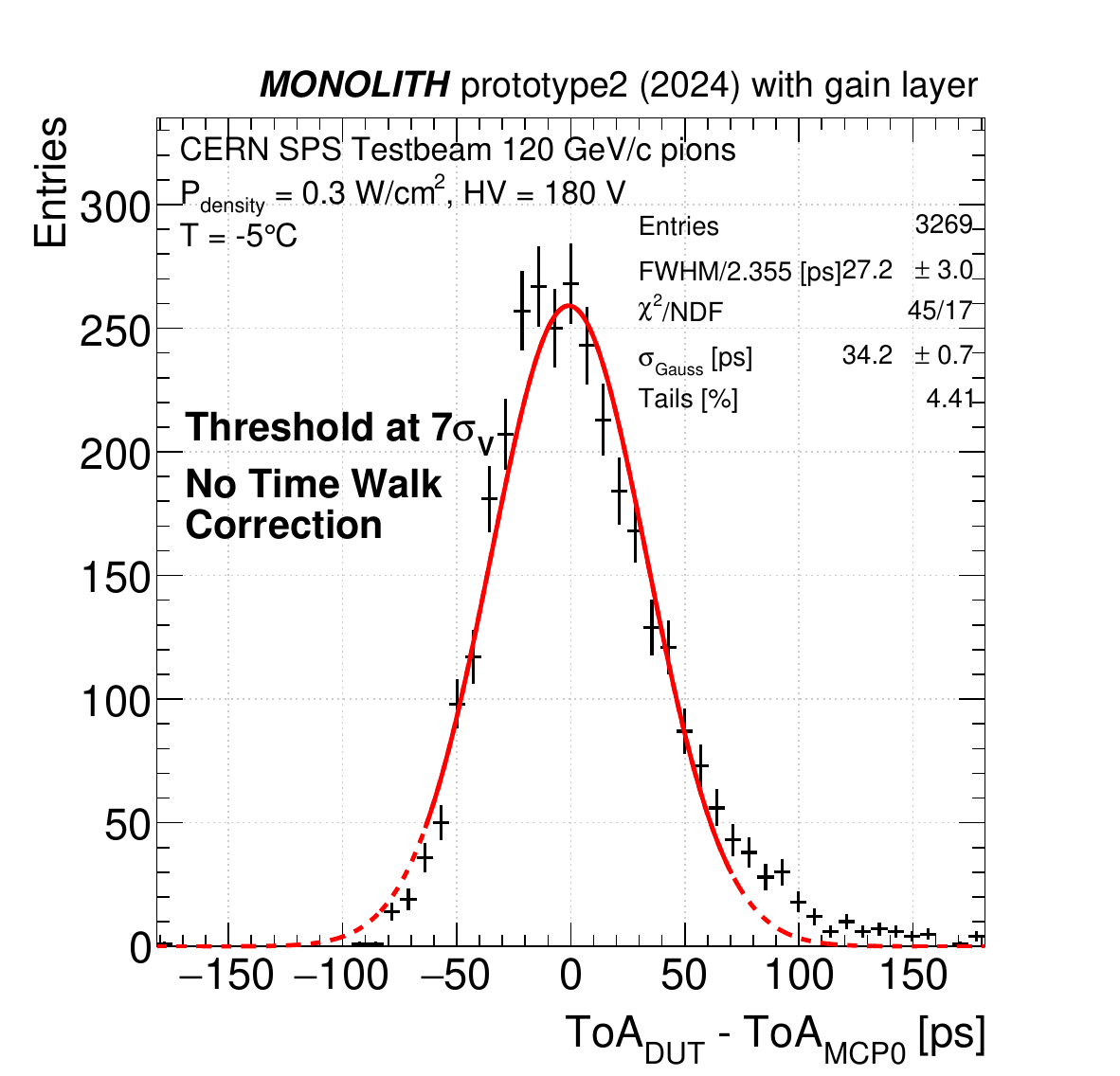}
\includegraphics[width=.32\textwidth]{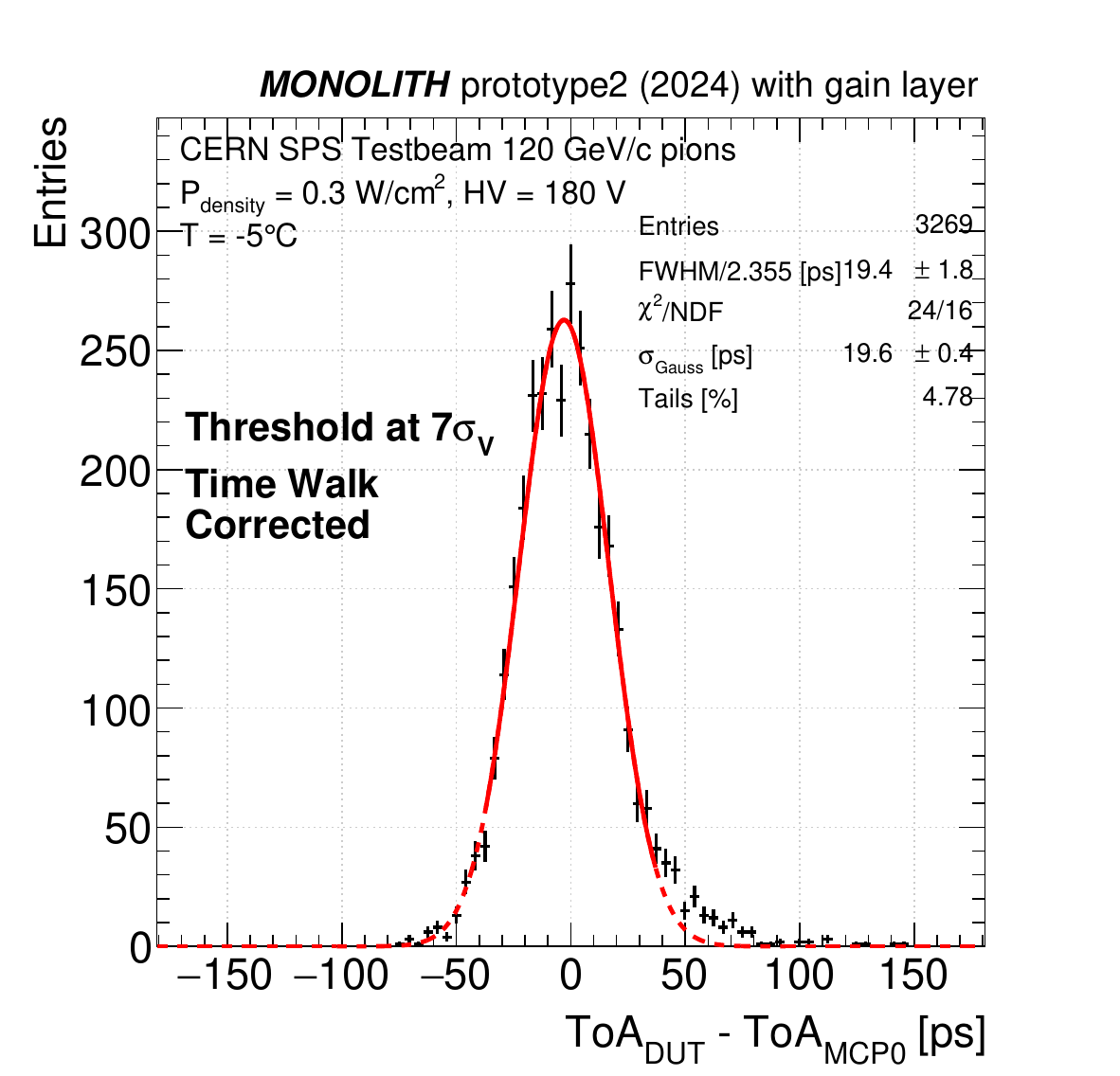}
\includegraphics[width=.32\textwidth]{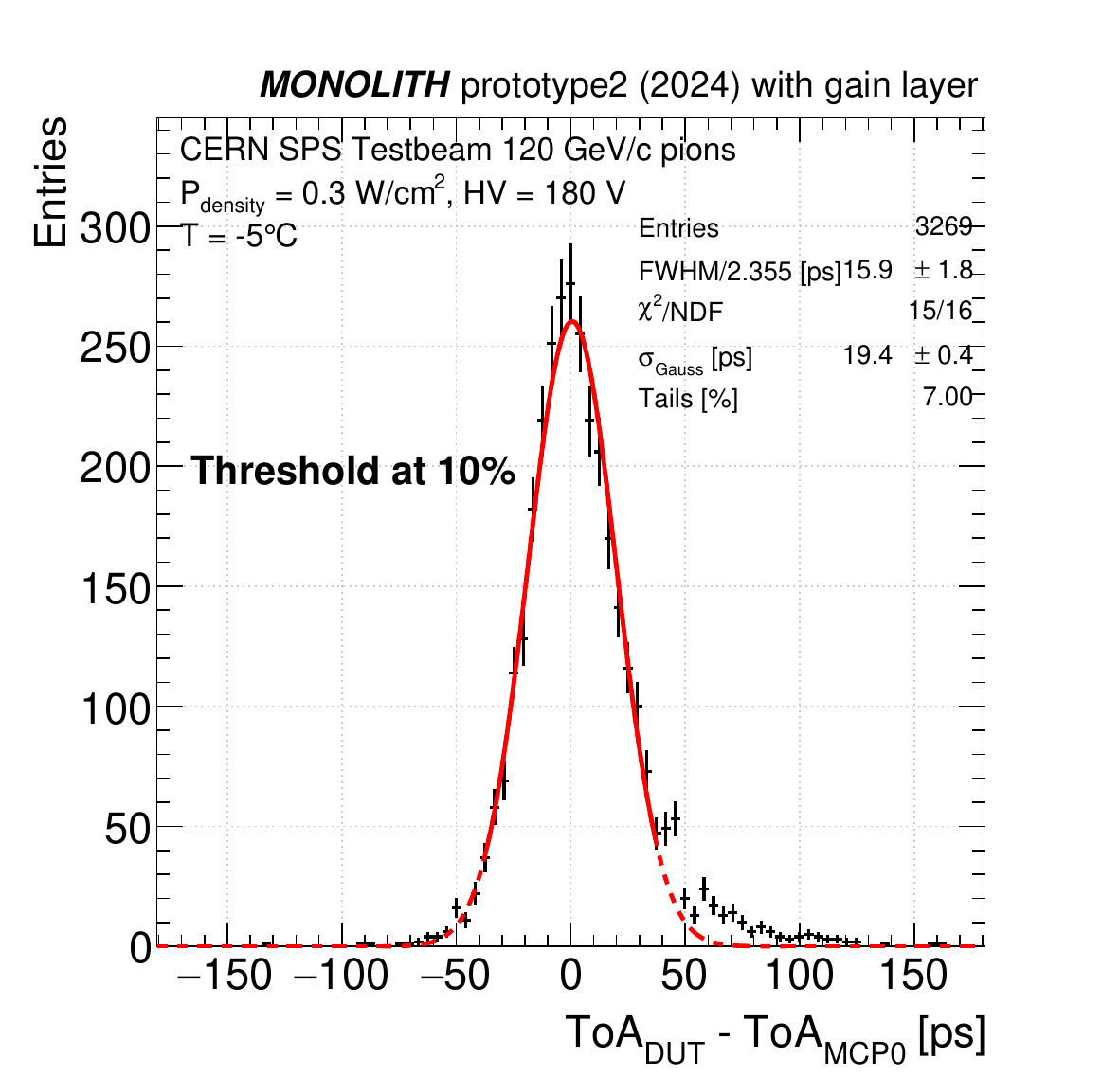}
\includegraphics[width=.32\textwidth]{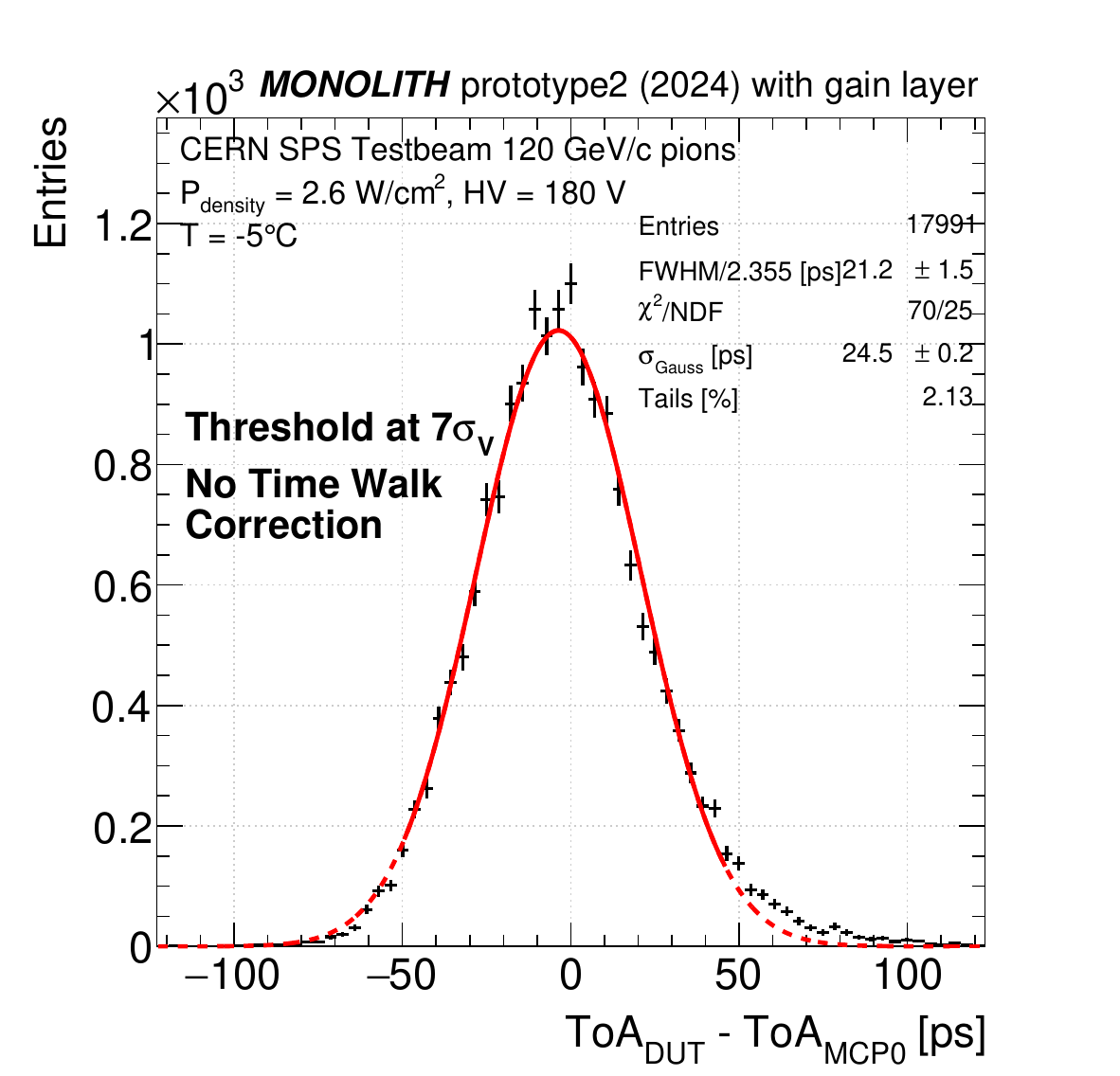}
\includegraphics[width=.32\textwidth]{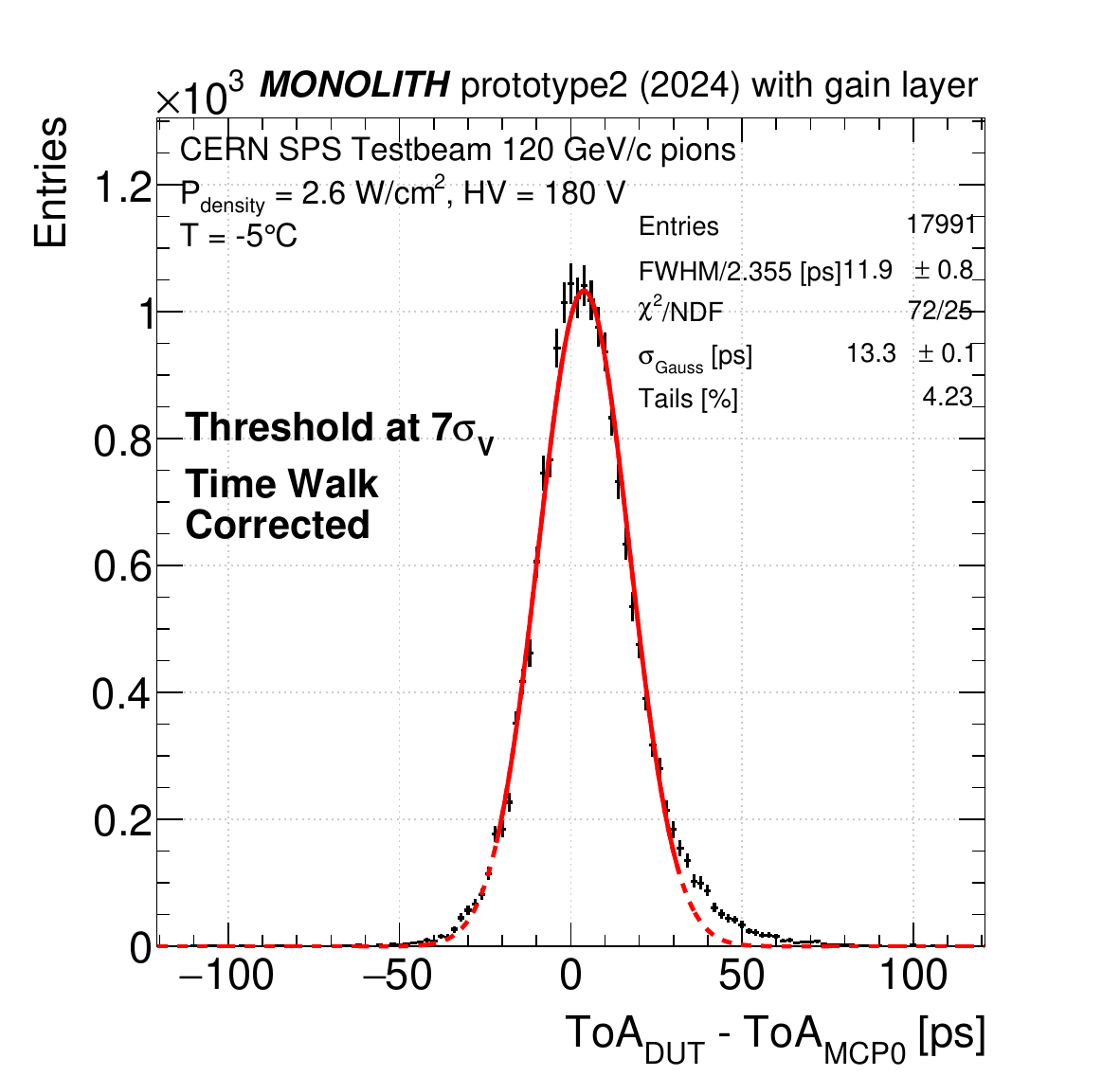}
\includegraphics[width=.32\textwidth]{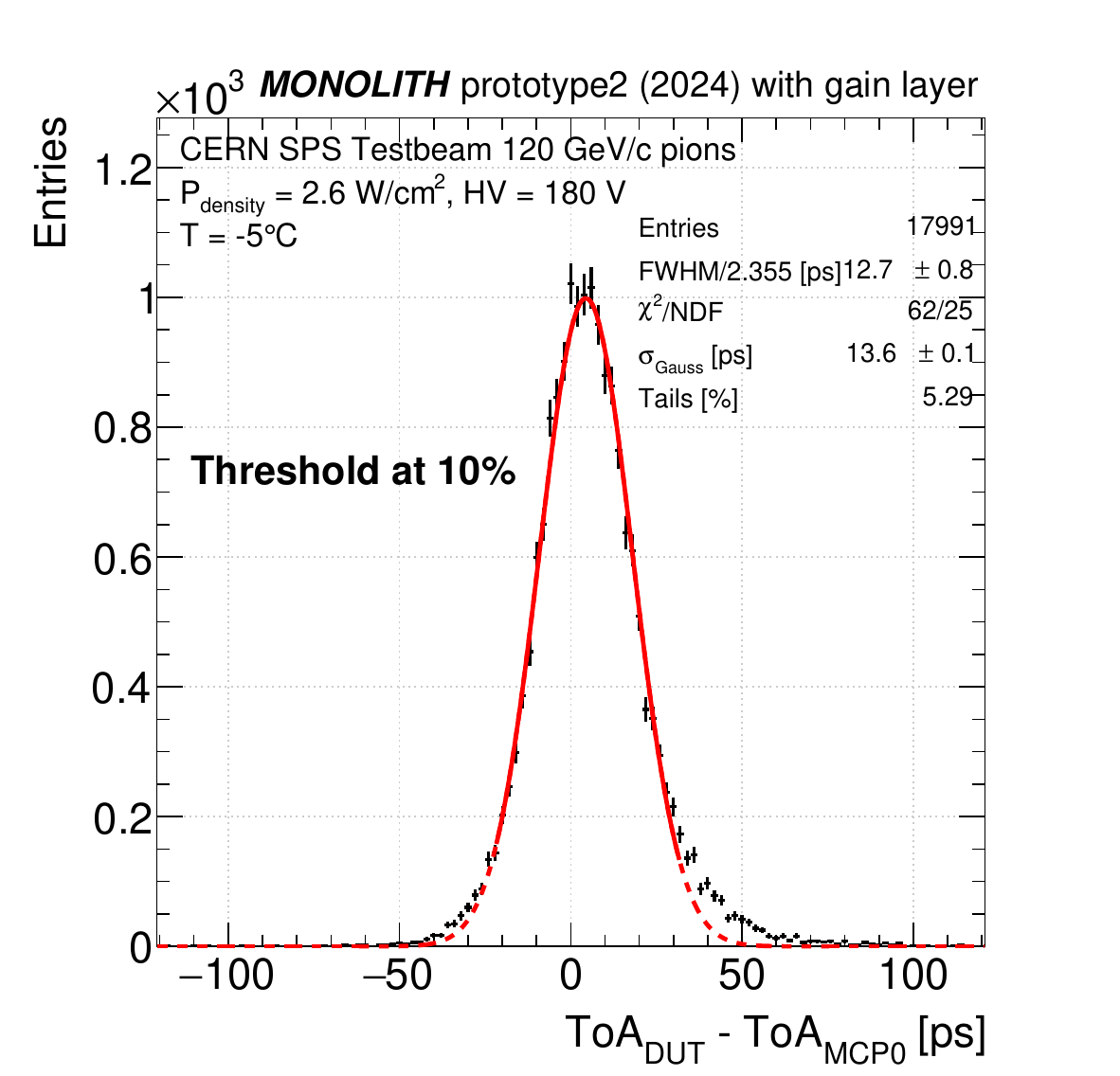}
\caption{\label{fig:ToA} 
ToA difference between the DUT and the MCP0 before (left panel) and after (center panel) time-walk correction when the DUT signal time is taken at a threshold V$_{\text{ToA}}$ = 7 $\sigma_V$.  
The right panels show the ToA difference obtained when the DUT signal time is taken at a threshold V$_{\text{ToA}}$ = 10\% of the signal height.
The two working points shown were taken at HV = 180 V and front-end power density of 0.3 (top) and 2.6  W/cm$^{2}$ (bottom). 
The distributions were fitted with a Gaussian function up to 20\% of the maximum.}
\end{figure}

The two panels on the left of Figure \ref{fig:ToA} 
show the ToA$_{\text{DUT}}$ - ToA$_{\text{MCP0}}$ distributions before the time-walk correction of the DUT signals for the datasets acquired at HV = 180 V with \pdensity~= 0.3 W/cm$^2$ (top) and 2.6 W/cm$^2$ (bottom). 
After subtraction of $\sigma_{t,\text{MCP0}}$, the time resolution results to be (33.2 $\pm$ 0.7) ps and (24.3 $\pm$ 0.2) ps, respectively.
Table~\ref{tab:summaryTiming} reports the time resolution before time-walk correction for all datasets with efficiency larger than 99\%.
As expected, the time resolution is measured to improve with increasing preamplifier power density and increasing sensor bias voltage.

\begin{figure}[htb!]
\centering 
\includegraphics[width=.48\textwidth]{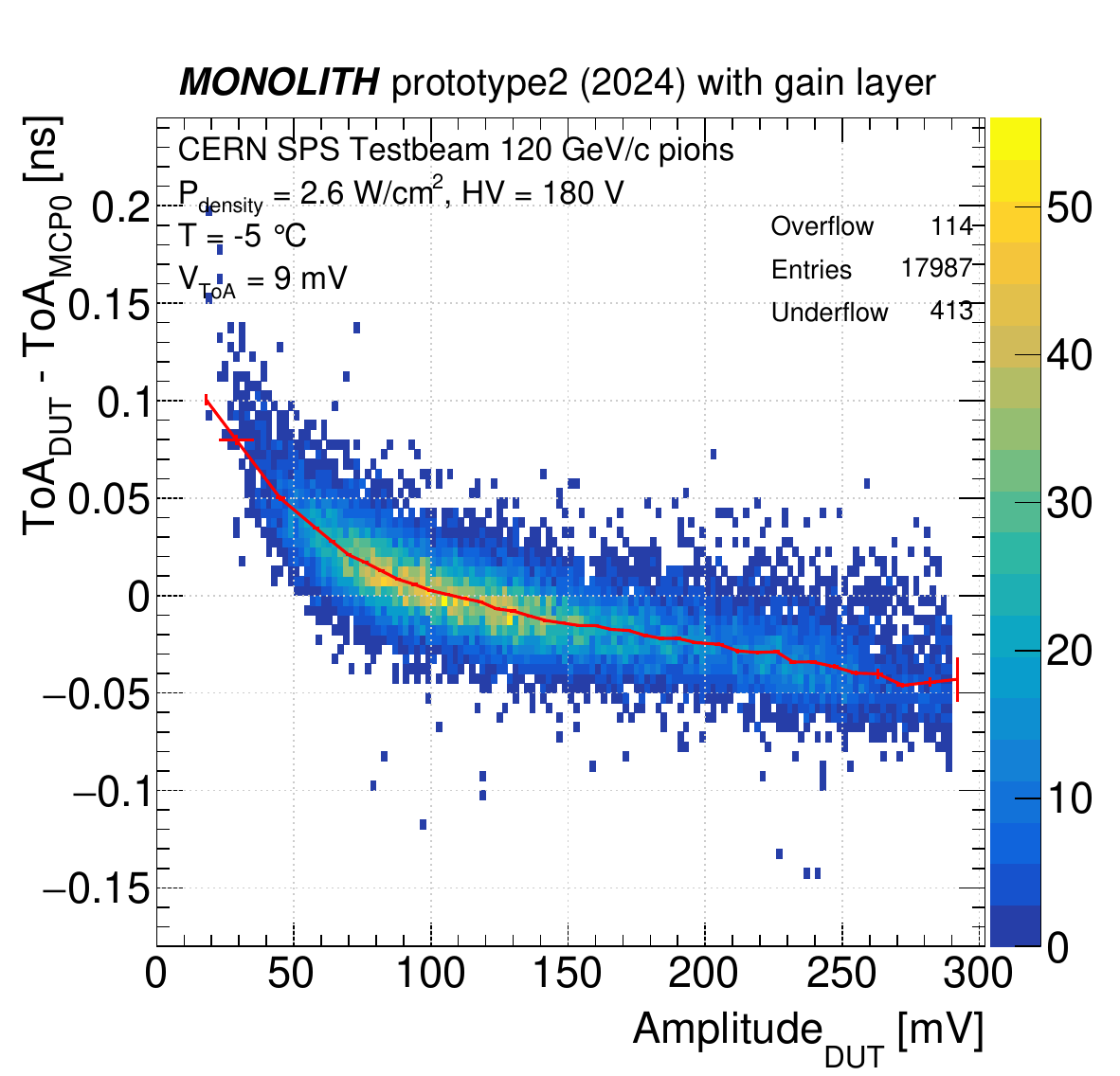}
\includegraphics[width=.48\textwidth]{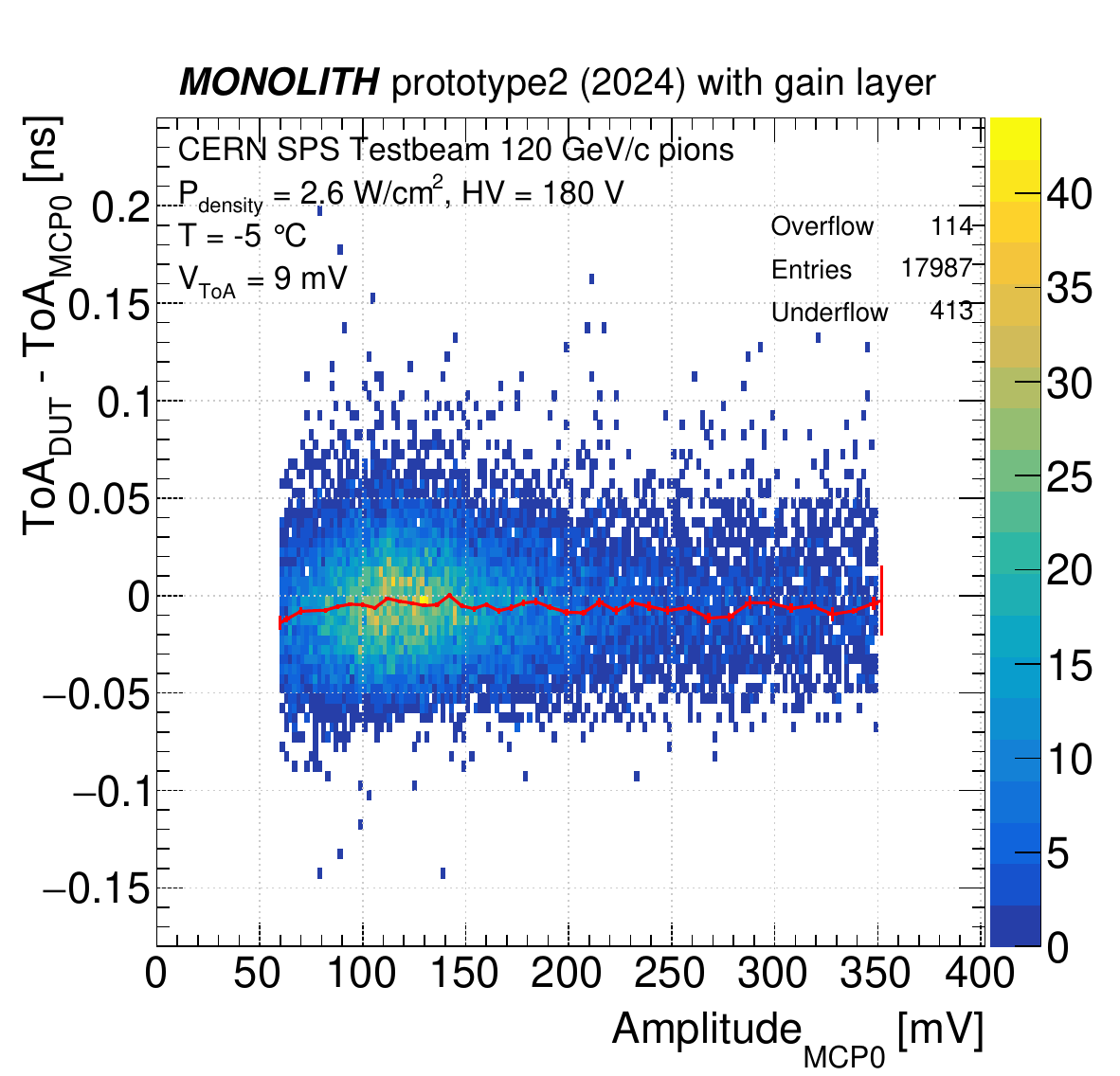}
\caption{\label{fig:TWcorrection} 
Difference between the times of arrival of the DUT and the MCP0 as a function of the amplitude of the DUT (left) or the MCP0 (right). The red crosses show the mean values of the difference of ToA, as obtained by a Gaussian fit in each vertical slice of the distributions. The red segments are the linear interpolations between the mean values.}
\end{figure}

The signals from the DUT were corrected for time walk by fitting the difference  
ToA$_{\text{DUT}}$ - ToA$_{\text{MCP0}}$	
within bins of the DUT signal amplitude using Gaussian functional forms. The mean values obtained from these fits were subsequently linearly interpolated to derive the time-walk correction.
The red line in the panel at the left of Figure~\ref{fig:TWcorrection} shows the time-walk correction function that was applied to the DUT in the case of the dataset taken with HV = 180 V and \pdensity~= 2.6 W/cm$^2$. The panel on the right of Figure~\ref{fig:TWcorrection} shows the ToA difference as a function of the amplitude of the MCP0 and confirms that the MCP0 did not need further time-walk correction.
The two panels at the center of Figure~\ref{fig:ToA} show the time-walk corrected ToA$_{\text{DUT}}$ - ToA$_{\text{MCP0}}$  distributions for the datasets acquired with HV = 180 V and \pdensity~= 0.3 and 2.6 W/cm$^2$.

\begin{figure}[htb!]
\centering 
\includegraphics[width=.6\textwidth]{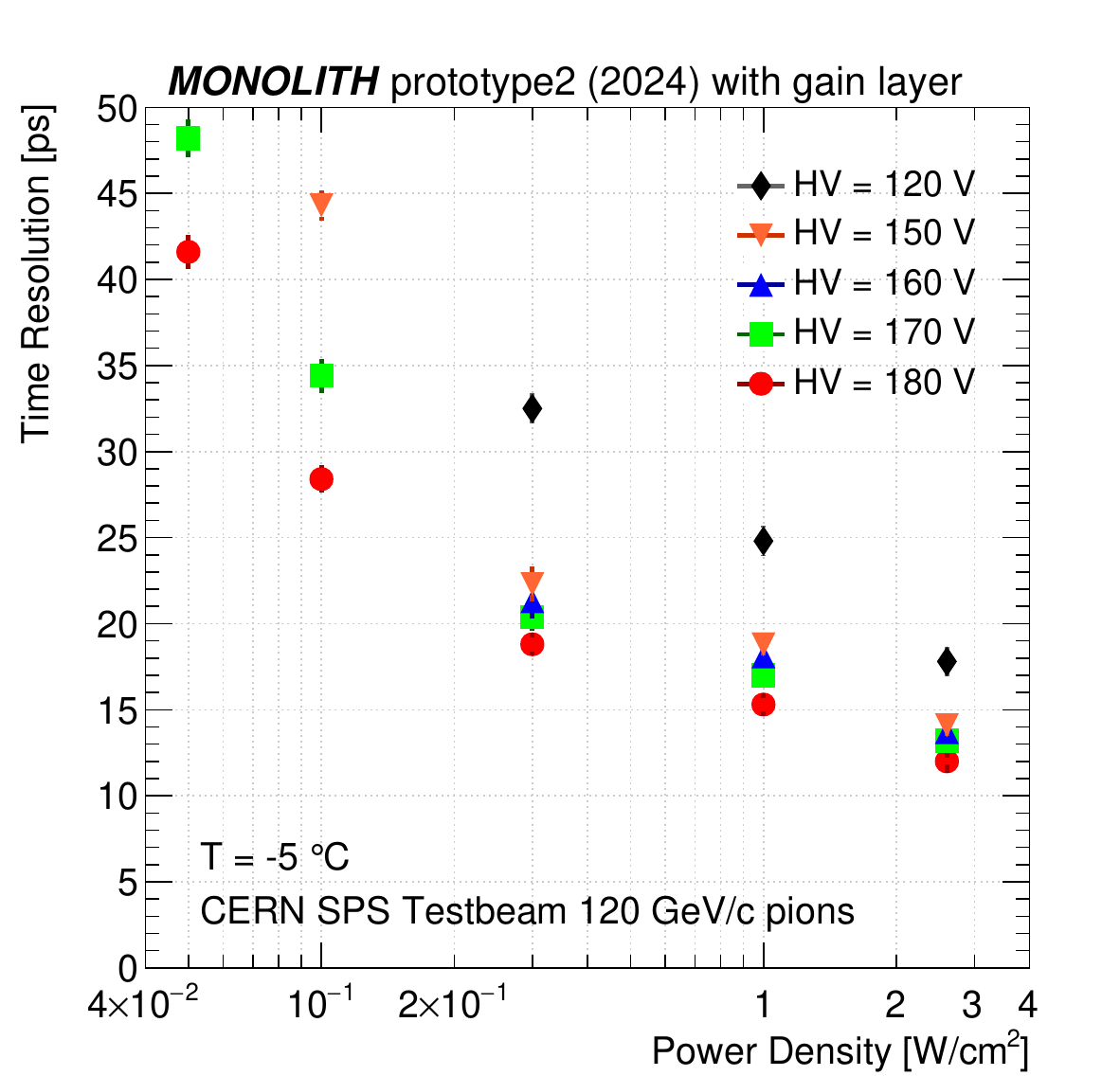}
\caption{\label{fig:timeResSummary} 
Time resolution as a function of the power density supplied to the preamplifier for five sensor bias voltages. 
Only the datasets in Table~\ref{tab:maintable} with operating conditions that provided detection efficiency larger than 99\% were considered for the measurement of the time resolution.
}
\end{figure}

The time-walk corrected time resolutions measured for all the acquired datasets with detection efficiency larger than 99\% are reported in Figure~\ref{fig:timeResSummary} and in Table~\ref{tab:summaryTiming}. 
In all cases, a time resolution better than 50 ps was measured.
At sensor bias voltage HV = 180 V (red dots in Figure~\ref{fig:timeResSummary}), the monolithic ASIC prototype with PicoAD sensor provides outstanding time resolution at low and high power density, namely 
$\sigma_{t, \text{DUT}}$ = (41.6 $\pm$ 1.0) ps at \pdensity~= 0.05 W/cm$^2$,
(18.8 $\pm$ 0.4) ps at 0.3 W/cm$^2$,
to reach (12.1 $\pm$ 0.3) ps at 2.6 W/cm$^2$.
Comparison of the time resolution measurements in Table~\ref{tab:summaryTiming}, obtained before and after time-walk correction, indicates that the correction significantly enhances the time resolution in all cases by approximately a factor of two.

Alternatively, the time resolution was also computed taking the time at 10\% of the maximum of the signal amplitude. The two right-side panels of Figure~\ref{fig:ToA} and the last column in Table~\ref{tab:summaryTiming} present the results obtained with this method.
This alternative method, which avoids the need for explicit time-walk correction, yields time resolutions nearly equivalent to those obtained after time-walk correction when using a fixed signal-voltage threshold. 

It should be noted that the ToA$_{\text{DUT}}$ - ToA$_{\text{MCP0}}$ distribution presents non-Gaussian tails at large positive values.
Through simulations, we verified that these delayed DUT signal times are caused by fluctuations in primary charge, which produce large clusters in the lower region of the absorption layer, positioned far from the collection electrode.
To account for the presence of these tails and evaluate their percentage,
the Gaussian fits were performed up to 20\% of the maximum height of the distribution. 
The fraction of non-Gaussian tails was then computed from the ratio between the total number of entries of the histogram and the integral of the  Gaussian function resulting from the fit.
In all datasets analysed, the tails were found to vary between 2\% and 10\%. Therefore, the time resolutions quoted here refer to 90\% or more of the recorded signals.


\begin{table}[]
\centering
\renewcommand{\arraystretch}{1.1}%
\begin{tabular}{|c|c|c|c|c|c|}
\hline

\vspace{-3pt}
~~~~HV~~~~~ & ~Electron~  & ~~~$P_{\it{density}}$~~~  & $\sigma_T$ (no TWC)    & ~~$\sigma_T$ (TWC)~~   & ~~~$\sigma_T$ (10\%)~~~  \\
{[}V] & Gain      & [W/cm$^2$]          & [ps]                      &  [ps]         & [ps]  \\
\hline
\multirow{3}{*}{120}& \multirow{3}{*}{10} 
                        & ~0.3   & $78.8 \pm 1.7$ & $32.5 \pm 0.7$ & $34.3 \pm 0.7$  \\
                    &   & ~\!1.0 & $50.9 \pm 0.5$ & $24.8 \pm 0.8$ & $26.8 \pm 0.9$  \\
                    &   & 2.6    & $42.0 \pm 0.5$ & $17.8 \pm 0.3$ & $19.4 \pm 0.2$  \\
\hline
\multirow{4}{*}{150}& \multirow{4}{*}{18}  
                        & ~0.1   & $98.1 \pm 2.1$ & $44.3 \pm 0.9$ & $45.9 \pm 0.8$  \\
                    &   & ~0.3   & $55.9 \pm 0.7$ & $22.3 \pm 0.3$ & $23.4 \pm 0.3$  \\
                    &   & ~\!1.0 & $36.9 \pm 0.8$ & $18.8 \pm 0.5$ & $20.0 \pm 0.4$  \\
                    &   & 2.6    & $32.5 \pm 0.3$ & $14.1 \pm 0.2$ & $14.8 \pm 0.3$  \\
\hline
\multirow{3}{*}{160}& \multirow{3}{*}{21}  
                        &   ~0.3 & $46.8 \pm 1.0$ & $21.3 \pm 1.0$ & $22.6 \pm 0.4$  \\
                    &   & ~\!1.0 & $35.1 \pm 0.8$ & $18.1 \pm 0.5$ & $18.4 \pm 0.4$  \\
                    &   &    2.6 & $29.5 \pm 0.7$ & $13.7 \pm 0.6$ & $14.8 \pm 0.6$  \\
\hline
\multirow{5}{*}{170}& \multirow{5}{*}{33} 
                        & ~~~0.05 & $88.6 \pm 1.6$ & $48.2 \pm 1.1$ & $47.1 \pm 0.7$  \\
                    &   &    ~0.1 & $65.2 \pm 1.3$ & $34.4 \pm 1.0$ & $33.7 \pm 0.6$  \\
                    &   &    ~0.3 & $38.1 \pm 1.2$ & $20.4 \pm 0.7$ & $20.4 \pm 0.7$  \\
                    &   &  ~\!1.0 & $29.9 \pm 0.5$ & $17.0 \pm 0.6$ & $17.4 \pm 0.5$  \\
                    &   &     2.6 & $24.9 \pm 0.6$ & $13.2 \pm 0.4$ & $13.4 \pm 0.4$  \\
\hline
\multirow{5}{*}{180}& \multirow{5}{*}{50} 
                        & ~~~0.05 & $73.8 \pm 1.4$ & $41.6 \pm 1.0$ & $40.9 \pm 0.7$  \\
                    &   &    ~0.1 & $52.4 \pm 1.2$ & $28.4 \pm 0.8$ & $28.2 \pm 0.6$  \\
                    &   &    ~0.3 & $33.2 \pm 0.7$ & $18.8 \pm 0.4$ & $18.6 \pm 0.4$  \\
                    &   &  ~\!1.0 & $26.9 \pm 0.6$ & $15.3 \pm 0.4$ & $15.8 \pm 0.4$  \\
                    &   &     2.6 & $24.3 \pm 0.2$ & $12.1 \pm 0.2$ & $12.5 \pm 0.2$  \\
\hline
\end{tabular}
\caption{
Electron gain and time resolution for the acquired datasets with a detection efficiency above 99\%. 
The time resolution was measured using several methods: fixed threshold without time-walk correction, fixed threshold with time-walk correction (used as the reference for the summary plots), and threshold at a constant fraction of the amplitude at 10\%.
}
\label{tab:summaryTiming}
\end{table}

\begin{figure}[htb!]
\centering
\includegraphics[width=.49\textwidth]{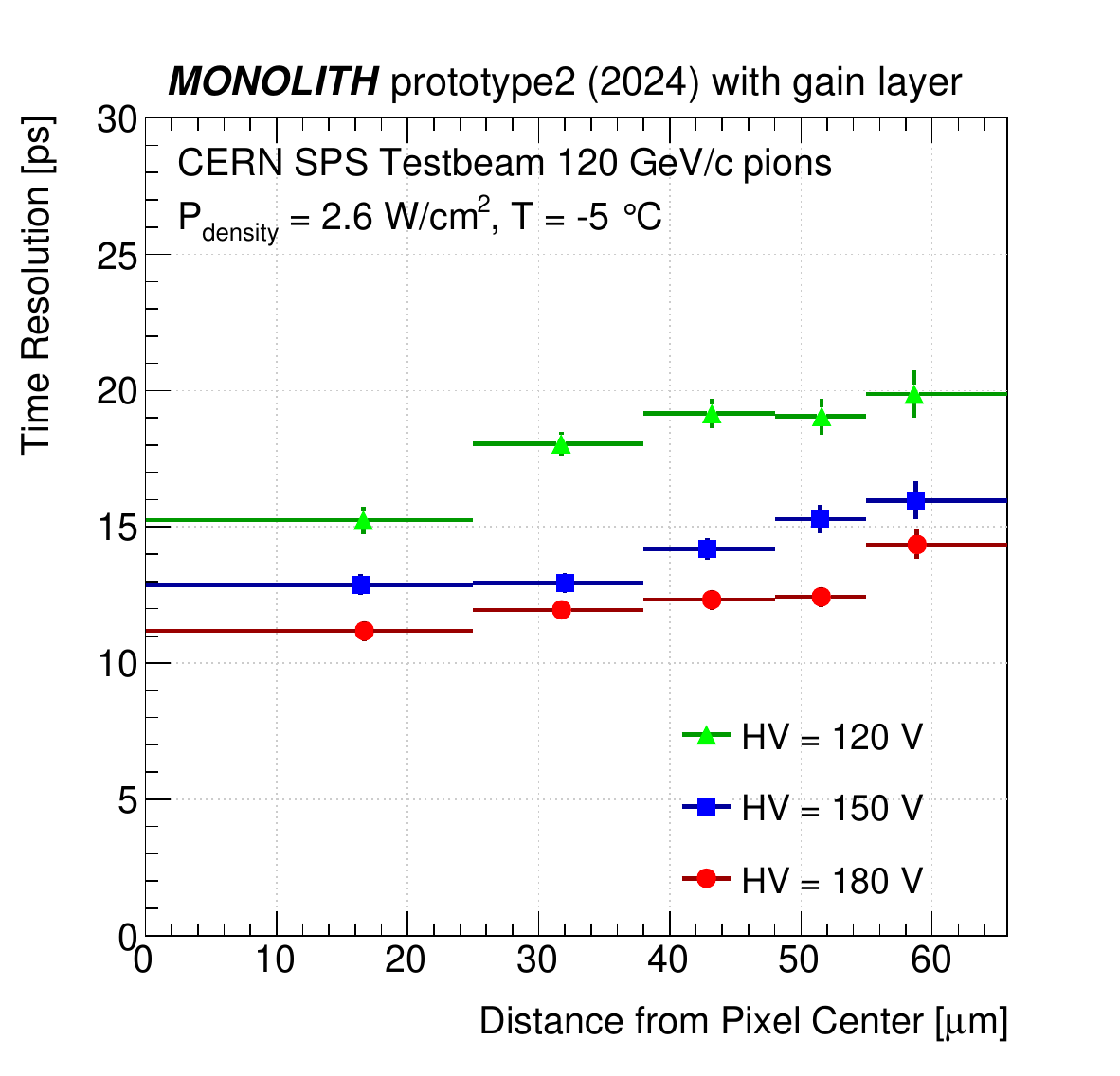}
\includegraphics[width=.49\textwidth]{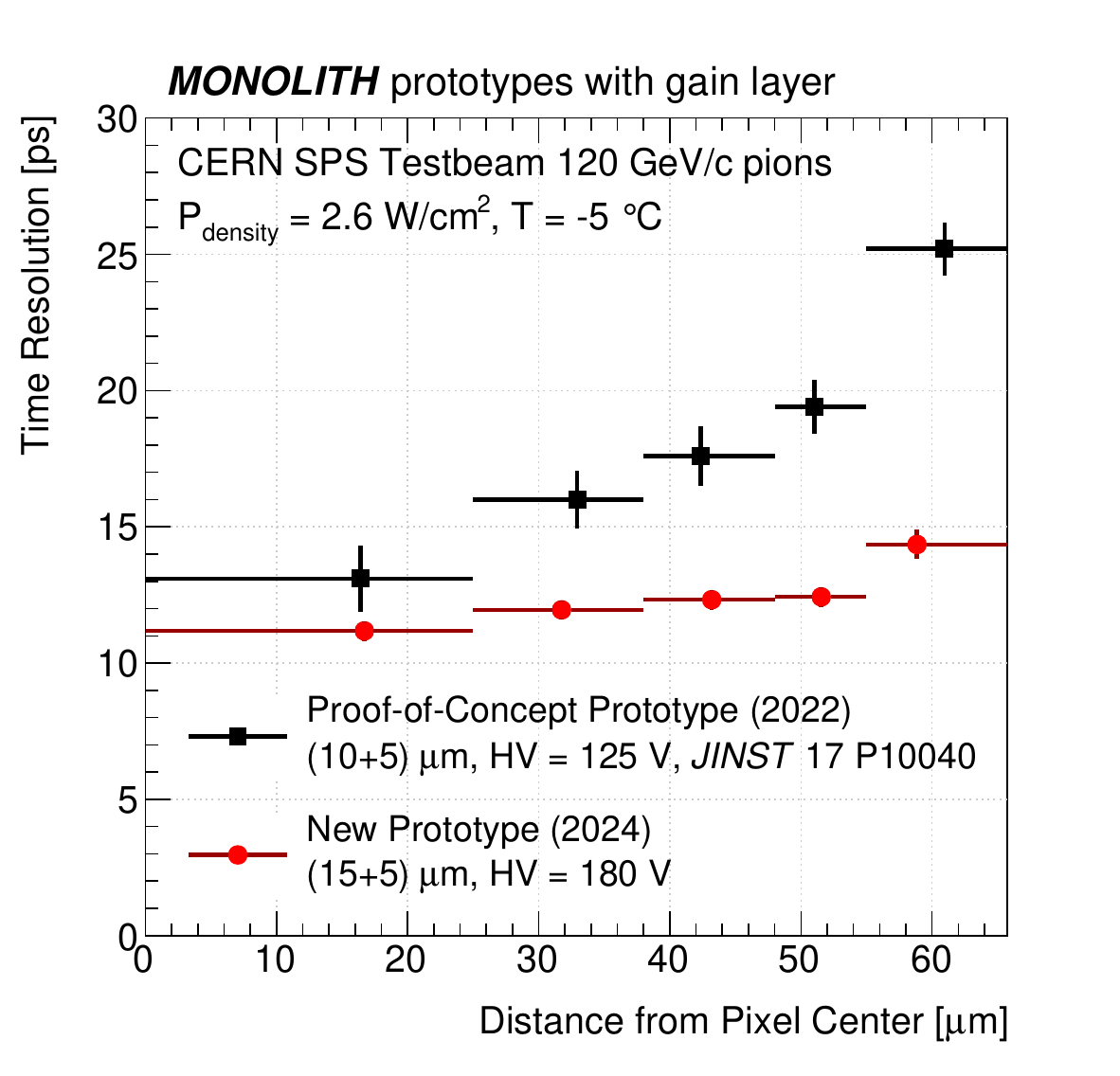}
\caption{\label{fig:Timeres_vs_distance}
Time resolution as a function of the distance from the pixel center. On the left is the scan for the three different sensor bias voltages. On the right is the comparison of the prototype studied in this paper and the proof-of-concept prototype. In each bin, the data points are plotted at the mean value of the track distance from the pixel center. The values from the proof-of-concept prototype are taken from~\cite{PicoAD_TB}.}
\end{figure}

The time resolution was analyzed in bins of the distance of the hit position from the pixel center.
The left panel in Figure \ref{fig:Timeres_vs_distance} shows the results obtained using five distance bins, defined to contain similar statistics, for the three acquired sensor bias voltage values at the highest \pdensity~= 2.6 W/cm$^2$.
The time resolution shows a worsening from the center to the edge of the pixel of approximately 3 ps for HV = 180 V and 5 ps for HV = 120 V, corresponding to approximately 30\%. 

The right panel of Figure~\ref{fig:Timeres_vs_distance} presents a comparison between the results of the current prototype and those of the PicoAD proof-of-concept prototype~\cite{PicoAD_TB}. In the proof-of-concept version, the average time resolution was 17 ps, with values ranging from 13 ps at the pixel center to 25 ps at the interpixel region, indicating a twofold worsening from center to edge. The reduced variation in time resolution with respect to the hit position in the current prototype is attributed to the decreased dependence of amplitude on the hit position, as shown in Figure~\ref{fig:ampRadius}.

The proof-of-concept prototype achieved a time resolution of 
$(17.3 \pm 0.4)$ ps
using a signal-processing technique referred to as "CFD-like" in~\cite{PicoAD_TB}. This method involved shifting the signal by 200 ps (half of its rise time), subtracting the shifted signal from the original, and determining the ToA at 25\% of the resulting signal amplitude.
This signal-processing approach was implemented to mitigate low-frequency noise components in that prototype, which originated from feedback injection through the power supply lines. Its adoption led to a significant improvement in time resolution compared to the $(24.1 \pm 0.4)$ ps~\cite{PicoAD_TB} achieved using an amplitude-based analysis similar to the method employed in this paper. In this updated prototype, where a redesigned front-end electronics system effectively resolved the feedback injection issue responsible for the low-frequency noise, the same method yields only a marginal improvement in time resolution of approximately 10\% and was consequently discontinued.

\section{Conclusions}
\label{sec:conclusions}

The detection efficiency and time resolution of a monolithic silicon detector produced in 2024 for the Horizon 2020 MONOLITH ERC Advanced project were measured using a 120 GeV/c pion beam. The prototype features a matrix of hexagonal pixels with a 100 $\mu$m pitch and front-end electronics that leverage the low noise and high-speed response of SiGe HBT technology. The ASIC incorporates the PicoAD sensor, which includes a continuous deep gain layer positioned between two thin epitaxial layers with 350 $\Omega$cm resistivity, resulting in a total sensor thickness of 20 $\mu$m.

Data were taken at sensor bias voltage between 90 and 180 V and front-end power densities between 0.05 and 2.6 W/cm$^2$. 
A fast oscilloscope was used to read the analog channels present in the pixel matrix and record the data.

At HV = 180 V, which is 20 V below the breakdown, and \pdensity~= 0.3 W/cm$^2$, the measured detection efficiency was ($99.92  ^{~\!+0.03}_{-0.10}$)\%.
When measuring the signal time at a fixed threshold of seven times the voltage noise, the time resolution was (33.2 $\pm$ 0.7) ps before time-walk correction and improved to (18.8 $\pm$ 0.4) ps after correction. 
Using the signal time at 10\% of the maximum amplitude, the time resolution remained at (18.6 $\pm$ 0.4) ps without the need for further time-walk correction.

At the same sensor bias voltage and at the highest front-end power used, \pdensity~= 2.6 W/cm$^2$, the detection efficiency achieved was ($99.99^{~\!+0.01}_{-0.02}$) \% and the time resolution after time-walk correction was (12.1 $\pm$ 0.3) ps.

All detection efficiency and time resolution values include the inter-pixel region. 

These results demonstrate that monolithic ASICs with  PicoAD sensor and SiGe HBT front-end electronics are outstanding candidates for a broad range of applications with power requirements spanning two orders of magnitude. They are well-suited for large-area applications, such as particle physics 4D trackers, where \pdensity~must stay well below 1 W/cm$^2$.
The PicoAD's full fill factor and ultra-thin design make it particularly suitable for applications requiring a fully active, low-material-budget timing detector.
It is also ideal for timing applications with fewer pixels, like direct-TOF LiDAR systems, where a power density of a few W/cm$^2$ is feasible.

\acknowledgments
This research is supported by the Horizon 2020 MONOLITH  ERC Advanced Grant ID: 884447. 
The authors wish to thank Coralie Husi, Javier Mesa, Gabriel Pelleriti, and the technical staff members of the University of Geneva and IHP Microelectronics.
Special thanks also go to Werner Riegler and Heinrich Schindler for their collaborative support and assistance with Garfield++ simulations. 
The authors also acknowledge EUROPRACTICE for providing design tools and MPW fabrication services, and the CERN SPS testbeam team for their valuable contributions.

\newpage
\bibliographystyle{unsrt}
\bibliography{bibliography.bib}
\end{document}